\documentclass[12pt]{article}

\usepackage{graphicx}
\usepackage{graphics}
\usepackage{amssymb}
\usepackage{amsmath}

\usepackage{epsf,amsfonts,hyperref}
\renewcommand{\appendix}[1]{
    \addtocounter{section}{1}
    \setcounter{equation}{0}
    \renewcommand{\thesection}{\Alph{section}}
    \section*{Appendix \thesection\protect\indent #1}
    \addcontentsline{toc}{section}{Appendix \thesection\ \ \ #1}
}
\newcommand\encadremath[1]{\vbox{\hrule\hbox{\vrule\kern8pt 
\vbox{\kern8pt \hbox{$\displaystyle #1$}\kern8pt} 
\kern8pt\vrule}\hrule}}
\def\enca#1{\vbox{\hrule\hbox{
\vrule\kern8pt\vbox{\kern8pt \hbox{$\displaystyle #1$}
\kern8pt} \kern8pt\vrule}\hrule}}

\newcommand\figureframex[3]{
\begin{figure}[bth]
\hrule\hbox{\vrule\kern8pt 
\vbox{\kern8pt \vbox{
\begin{center}
{\mbox{\epsfxsize=#1.truecm\epsfbox{#2}}}
\end{center}
\caption{#3}
}\kern8pt} 
\kern8pt\vrule}\hrule
\end{figure}
}
\newcommand\figureframey[3]{
\begin{figure}[bth]
\hrule\hbox{\vrule\kern8pt 
\vbox{\kern8pt \vbox{
\begin{center}
{\mbox{\epsfysize=#1.truecm\epsfbox{#2}}}
\end{center}
\caption{#3}
}\kern8pt} 
\kern8pt\vrule}\hrule
\end{figure}
}

\makeatletter
\@addtoreset{equation}{section}
\makeatother
\newtheorem{theorem}{Theorem}[section]

\newtheorem{remark}{Remark}[section]
\newtheorem{proposition}{Proposition}[section]
\newtheorem{lemma}{Lemma}[section]
\newtheorem{corollary}{Corollary}[section]
\newtheorem{definition}{Definition}[section]
\def\br{\begin{remark}\rm\small}
\def\er{\end{remark}}
\def\bt{\begin{theorem}}
\def\et{\end{theorem}}
\def\bd{\begin{definition}}
\def\ed{\end{definition}}
\def\bp{\begin{proposition}}
\def\ep{\end{proposition}}
\def\bl{\begin{lemma}}
\def\el{\end{lemma}}
\def\bc{\begin{corollary}}
\def\ec{\end{corollary}}
\def\beaq{\begin{eqnarray}}
\def\eeaq{\end{eqnarray}}
\newcommand{\proof}[1]{{\noindent \bf proof:}\par
{#1} $\square$}

\newcommand{\beq}{\begin{equation}}
\newcommand{\eeq}{\end{equation}}
\newcommand{\bea}{\begin{eqnarray}}
\newcommand{\eea}{\end{eqnarray}}

\renewcommand{\and}{{\qquad {\rm and} \qquad}}

\newcommand{\Tr}{\operatorname{Tr}}

\newcommand{\Res}{\mathop{\,\rm Res\,}}
\newcommand{\order}{\operatorname{order}} 
 
\newcommand{\supp}{\operatorname{supp}} 

\newcommand{\CC}{{\mathbb C}}
\newcommand{\ZZ}{{\mathbb Z}}
\newcommand{\RR}{{\mathbb R}}

\newcommand{\td}[1]{{\tilde{#1}}}

\newcommand{\om}{\omega}

\newcommand{\ii}{{\mathrm{i}}}
\newcommand{\e}{{\,\rm e}\,}

\newcommand{\Pint}{{\int\kern -1.em -\kern-.25em}}

\renewcommand{\Im}{{\mathrm{Im}}}

\newcommand{\curve}{{\Sigma}}

\newcommand{\Ker}{\operatorname{Ker}}

\newcommand{\spcurve}{{\mathcal S}}
\newcommand{\curverond}{\underline{\curve}}
\newcommand{\genus}{{\mathfrak g}}

\newcommand{\acycle}{{\cal A}}
\newcommand{\bcycle}{{\cal B}}

\newcommand{\y}{{\rm{y}}}
\newcommand{\x}{{\rm{x}}}

\begin{document}
\sloppy


\pagestyle{empty}
\hfill IPhT-T23/??? 
\addtolength{\baselineskip}{0.20\baselineskip}
\begin{center}
\vspace{26pt}
{\large \bf {Generalized cycles on Spectral Curves}}
\newline
\vspace{26pt}

{\sl B.\ Eynard}$^{1}$\hspace*{0.05cm}\\
\vspace{6pt}
$^1$
\ Institut de physique th\'eorique, Universit\'e Paris Saclay, 
\\
CEA, CNRS, F-91191 Gif-sur-Yvette, France
\end{center}

\vspace{20pt}
\begin{center}
{\bf Abstract}

Generalized cycles can be thought of as the extension of form-cycle duality between holomorphic forms and cycles, to meromorphic forms and generalized cycles.

They appeared as an ubiquitous tool in the study of spectral curves and integrable systems in the topological recursion approach. They parametrize deformations, implementing the special geometry, where moduli are periods, and derivatives with respect to moduli are other periods, or more generally "integrals", whence the name "generalized cycles".
They appeared over the years in various works, each time in specific applied frameworks, and here we provide a comprehensive self-contained corpus of definitions and properties for a very general setting. The geometry of generalized cycles is also fascinating by itself.

\end{center}
%





\vspace{26pt}
\pagestyle{plain}
\setcounter{page}{1}



\section{Introduction}

The idea of generalized cycles, is to extend to meromorphic 1-forms the \textbf{form-cycle duality} that exists for {\em{holomorphic}} 1-forms with {\em {cycles}}.
Generalized cycles could thus be called "{\em{meromorphic cycles}}".
They were introduced because of their ubiquitous role in the theory of Topological Recursion \cite{EO07}, in particular they parametrize deformations of spectral curves, they implement some "special geometry" relations, i.e. the fact that derivatives with respect to some periods of the spectral curve, are also periods (integrals on the curve). 

\medskip \textbf{Motivation example 1 : Seiberg-Witten theory}

In Seiberg Witten theory, there is a compact Riemann surface $\curve$ of some genus $\genus$ and a 1-form $\y$. 
The curve $\curve$ is endowed with a choice of $2\genus$ cycles (in the usual sense) $\acycle_1,\dots,\acycle_{\genus},\bcycle_1,\dots,\bcycle_{\genus}$, with symplectic intersections:
\beq
\acycle_i\cap \bcycle_j = \delta_{i,j}
\quad , \quad
\acycle_i\cap \acycle_j = 0
\quad , \quad
\bcycle_i\cap \bcycle_j = 0.
\eeq
Such a choice,  not unique, is called a "Torelli marking of cycles".
The $\acycle$-periods of $\y$
\beq
a_i = \frac{1}{2\pi\ii} \oint_{\acycle_i} \y
\eeq
are moduli of the Seiberg-Witten curve.
The Seiberg-Witten prepotential is defined as a function $F_0$ of the moduli that satisfies "special geometry":
\beq
\frac{\partial}{\partial a_i} F_0 = \oint_{\bcycle_i} \y.
\eeq
In other words, derivatives with respect to $\acycle$ periods are $\bcycle$ periods.

\smallskip
\textbf{Example 2: random matrices}. Consider a random Hermitian matrix $M\in H_N$ with probability law of the form
\beq
\label{eqdef:Z1MM}
\frac{1}{Z} e^{-N\Tr V(M)} D_{\text{Lebesgue}}(M) 
\qquad , \quad 
Z = \int_{H_N}e^{-N\Tr V(M)} D_{\text{Lebesgue}}(M) ,
\eeq
where $D_{\text{Lebesgue}}$ is the Lebesgue measure of $H_N$, and  $Z$ is the partition function (that makes the total probability equal to 1), and $V(M)=\sum_{k=1}^d \frac{t_k}{k} M^k$ is a polynomial called the potential.
The expectation value of the resolvant is written
\beq
W_1(x) = \mathbb E\left( \frac{1}{N}\Tr (x-M)^{-1} \right)
= \sum_{k=0}^\infty \frac{1}{x^{k+1}} \ \mathbb E\left( \frac{1}{N}\Tr M^k \right).
\eeq
Let us denote the 1-form
\beq
\y = V'(x)dx-W_1(x)dx
\eeq
which is a 1-form analytic outside of the support of eigenvalues distribution.
Since $\y = V'(x)dx + O(1/x)dx$ at large $x$, the coefficients of $V(x)$, called here the "moduli" are worth
\beq
t_k = -\Res_{\infty} x^{-k} \y .
\eeq
And taking the derivative with respect to  $t_k$ in \eqref{eqdef:Z1MM} gives
\beq
\frac{\partial}{\partial t_k} \frac{1}{N^2}\ln Z = \mathbb E\left( \frac{-1}{kN}\Tr M^k \right) =\frac{1}{k}\Res_{\infty} x^k W_1(x)dx = \frac{-1}{k}\Res_{\infty} x^k \y.
\eeq
This is a special case of the Miwa--Jimbo relation for Tau--functions in isomonodromic integrable systems.
But this is another example of special geometry, where the derivative with respect to an integral of $\y$ (a residue), is another integral of $\y$ (another residue).

Moreover, it is known in random matrix theory, that in the large $N$ limit, $W_1(x)$ tends to an algebraic function of $x$ and thus $\y$ tends to a meromorphic 1-form on a Riemann surface $\curve$. 
Topological recursion was invented to compute the asymptotic expansion at large $N$, and to satisfy the Miwa-Jimbo relation to all orders in the expansion. 

\medskip
The same kind of relations appear in many other integrable systems, except that instead of integrals over closed cycles, one can have other types of integrals: integrals on open chains, or residues...
on some Riemann surface $\curve$ equipped with a certain differential 1-form $\y$.
This is what led, in the topological recursion framework  to the notion of generalized cycles associated to a spectral curve.

\section{Spectral Curve}

Generalized cycles are associated to spectral curves.

\subsection{Definitions}

\bd[Spectral curve]
A spectral curve $\spcurve=(\curve,\x,\y,B)$ is the data of
\begin{itemize}
\item $\curve$ is an open Riemann surface, not necessarily compact nor connected.
\item $\x:\curve\to \curverond$ is a ramified covering of a Riemann surface $\curverond$ called the "base", and most of the time we shall choose $\curverond=\CC P^1$.
\item  $\y$ is a meromorphic 1-form on $\curve$
 (in any chart $\y=f(\zeta)d\zeta$ where $f(\zeta)$ is a meromorphic function of a local coordinate $\zeta$), 
\item $B$ is a symmetric $1\boxtimes 1$-form with a double pole on the diagonal, with bi-residue 1:
\beq
B\in H^0(\curve\times\curve, K_\curve\overset{\text{sym}}\boxtimes K_\curve(2\text{diag}) )_1
\eeq
which means that in in any pair of charts $U_i\times U_j$, and local coordinates $\zeta_i,\zeta_j$ (and we use the same charts and coordinates for neighborhoods of the diagonal), $B$ behaves as
\beq
B(z_1,z_2) \sim \frac{\delta_{i,j}}{(\zeta_i(z_1)-\zeta_j(z_2))^2}d\zeta_i(z_1)\otimes d\zeta_j(z_2) + \text{bi-holomorphic}.
\eeq
\end{itemize}
\ed
We recall the usual examples of $B$ in section \ref{sec:exB} below.


\br
If $\curverond = \CC P^1$, it is customary to write 
\beq
\y = yd\x
\eeq 
where $y$ is a meromorphic function on $\curve$.
In other words one can identify $T^*\CC P^1$ with $\CC\times \CC P^1 $.
\er

The following is an immediate consequence of the definition
\bp[Lagrangian immersion]
The map $\ii:\curve \to T^*\curverond$ given by $\ii(p)=(\x(p),\y(p))$ is a Lagragian immersion of $\curve$ into the cotangent space of the base curve
\bea
\curve & \overset{\ii=(\x,\y)}{\hookrightarrow} & T^*\curverond \cr
\x & \searrow  & \downarrow \cr
& & \curve
\eea
$\y$  is the pullback by $\ii$ of the tautological 1-form of $T^*\curverond$ i.e. the Liouville form.
\ep


\bd[Meromorphic forms]
\label{def:M1forms}

Here we define a meromorphic 1-form on $\curve$ to be a 1-form holomorphic on the interior of $\curve\setminus\text{isolated points}$, and bounded by a rational function in a local coordinate in any neighborhood, and in fact a meromorphic 1-form is the equivalence class of such 1-forms that differ on at most a set of isolated points:
\beq
\omega_1 \equiv \omega_2 \quad
\text{iff they are equal on } \curve \setminus \{\text{isolated points}\}.
\eeq
We shall call pole, any point where a 1-form is not holomorphic.

Let $\mathfrak M^1(\curve)$ the sheaf of  meromorphic 1-forms.
Notice that it is of infinite uncountable  dimension.
\ed

We needed to adapt the residue definition in case we would choose two representatives that differ exactly at the point where we compute the residue, and with the following definition the residue depends only on the equivalence class:
\bd[Residue]
The residue of a 1-form is defined as 
\beq
\Res_p \omega = \frac{1}{2\pi\ii}\liminf_{r\to 0} \oint_{\mathcal C_{p,r}} \omega
\eeq
where in a given chart around $p$, $\mathcal C_{p,r}$ is a circle of radius $r>0$. The projective limit $r\to 0$ is independent of a choice of representative of $\omega$ and is  well defined because for any representative the integral is constant as soon as $r$ is smaller than the distance of the closest non holomorphic isolated point to $p$.
\ed

\subsection{The bidifferential \texorpdfstring{$B$}{Lg}}
\label{sec:exB}

The bidifferential $B$ is here quite arbitrary.
The reader can skip this subsection whose role is only to illustrate the zoology of possible $B$ in some specific examples, and relies on a prior knowledge of {\em{Theta functions}} on Riemann surfaces. For this subsection we refer to classical textbooks on compact Riemann surfaces as \cite{farkas2012riemann,fay1973theta,TataLectures}.

\medskip

Notice that one can add to $B$ any symmetric tensor product of holomorphic forms, so that the space of possible $B$ is an affine space:
\beq
B + \Omega^1(\curve)\overset{\text{sym}}\boxtimes \Omega^1(\curve).
\eeq

Let us consider the particularly interesting case where $\curve$ is a connected compact Riemann surface of some genus $\genus$.

\begin{itemize}

\item \textbf{Case genus =0}, i.e. $\curve\sim \CC P^1$ the Riemann sphere. In that case there is no non-zero holomorphic 1-form, and $B$ is unique:
\beq
B(z_1,z_2) =  \frac{1}{(z_1-z_2)^2} dz_1 \otimes dz_2
= d_{z_1}\otimes d_{z_2} \left(\ln{(z_1-z_2)}\right)
\eeq
where $z$ is the canonical coordinate of $\CC P^1 = \CC \cup \{\infty\}$.

\item \textbf{Case genus =1}, i.e. $\curve\sim T_\tau = \CC/(\ZZ+\tau\ZZ) $ the torus of modulus $\tau$, with its canonical coordinate $z\equiv z+1\equiv z+\tau$.
In that case $B$ is of the form
\beq
B(z_1,z_2) =  \left( \wp(z_1-z_2) + C\right) dz_1 \otimes dz_2
\eeq
where $\wp$ is the Weierstrass function, and $C$ is an arbitrary constant that parametrizes the one-dimensional affine space of possible $B$.

Some choices of $C$ are particularly interesting:
\begin{itemize}
\item $C=G_2(\tau)$ the second Eisenstein series. In that case $B$ is called the \textbf{Bergman kernel} (cf \cite{BergSchif, Kokotov2009OC2}). It is normalized on the cycle $\acycle = [\frac\ii{2},1+\frac{\ii}{2}]$:
\beq
\forall z_2 , \quad \int_{z_1\in \acycle} B(z_1,z_2) = 0.
\eeq
It is the second derivative of the log of the Theta function
\beq
B(z_1,z_2) = d_{z_1}\otimes d_{z_2} \ \ln\Theta(z_1-z_2+\frac12+\frac12\tau )
\eeq
where $\Theta=\vartheta_3 $ is the Riemann Theta-function, or also the 3rd Jacobi Theta function $\vartheta_3$.

\item $C=G_2(\tau) + \pi/\Im\tau$. In that case $B$ is called the \textbf{Schiffer kernel} (cf \cite{BergSchif, Kokotov2009OC2}). It is related to the \textbf{Green function}:
\beq
B(z_1,z_2) = d_{z_1}\otimes dz_{2} \ \mathcal G(z_1,z_2).
\eeq
where $\mathcal G(z_1,z_2):\curve\times\curve \to \RR$ has a logarithmic singularity on the diagonal, and satisfies the Poisson equation $\Delta_{z_1} \mathcal G(z_1,z_2) = 4\pi \delta^{(2)}_{z_2}(z_1)$.

\item Let $\zeta\in \CC$, and choose $C=G_2(\tau)-\frac{d^2}{d\zeta^2}\ln\Theta(\zeta+\frac12+\frac12\tau ) $. In that case $B$ is called the \textbf{Klein kernel} (cf \cite{Kokotov2009OC2}).

\end{itemize}

\item \textbf{Higher genus:}
we have
\bea
B(z_1,z_2) 
&=& d_{z_1}\otimes d_{z_2} \ \ln\Theta\left(\mathfrak a(z_1)-\mathfrak a(z_2) + \frac12 (\mathbf n + \tau \mathbf m) \right) \cr
&& + 2\pi\ii \sum_{i=1}^\genus\sum_{j=1}^\genus \kappa_{i,j} \omega_i(z_1)\otimes \omega_j(z_2)
\eea
where $\Theta:\CC^{\genus}\to\CC $ is the higher genus Riemann Theta function, $\mathfrak a:\curve\to\CC^{\genus} $ is the Abel map, and $\mathbf n,\mathbf m \in \ZZ^{\genus}\times \ZZ^{\genus} $ is an integer non-singular characteristic, odd $(\mathbf n,\mathbf m)\in 2\ZZ+1$, and $\omega_{i=1,\dots,\genus}$ is a basis of holomorphic 1-forms such that $d\mathfrak a=(\omega_1,\dots,\omega_{\genus})$.

The symmetric $\genus\times \genus$ matrix $\kappa$ parametrizes  the affine space of possible $B$.

$\kappa=0$ is the \textbf{Bergman kernel}.

$\kappa=(\bar\tau -\tau)^{-1}$ is the \textbf{Schiffer kernel}.

\item There exist also Klein kernel and many others, see \cite{fay1973theta,eynard2018notes}.

\item \textbf{Non compact curves:} In that case we have a much larger choice of possible $B$, in an infinite dimensional affine space.
In particular if $\curve$ is an open subdomain of a compact curve, one can choose the restriction of the Bergman kernel.
But many other choices have been considered.

\end{itemize}

\section{Generalized cycles}

Since we want to define a duality pairing between meromorphic forms and "generalized cycles", we have to start by taking "generalized cycles" in the algebraic dual of $\mathfrak M^1(\curve)$.
The algebraic dual is a very nasty space,  we shall consider a subspace much nicer.

\subsection{Definitions}

Let $\mathfrak M^1(\curve)^*$ the algebraic dual of $\mathfrak M^1(\curve)$. i.e. the complex vector space of all linear forms on $\mathfrak M^1(\curve)$.

\bd[Notation Poincaré pairing]
If $\gamma\in \mathfrak M^1(\curve)^*$ and $\omega\in \mathfrak M^1(\curve)$, we will often denote the duality pairing with the same symbol as an integral:
\beq
\gamma(\omega)  \overset{\text{notation}}{=} <\gamma,\omega> \overset{\text{notation}}{=} \int_\gamma \omega.
\eeq

\ed

The bidifferential $B$ is, in each chart$\times$chart, a meromorphic form  valued in meromorphic forms. 
In other words, for $p\in\curve$, in particular in a chart $U$ with local coordinate $\xi$, then $B(.,p)/d\xi(p)$ is a meromorphic 1-form on $\curve$, that has a double pole at $p$ and no other pole.
We shall define $<\gamma,B>$ as:
\beq
p \mapsto   <\gamma,B(.,p)/d\xi(p)> d\xi(p) .
\eeq
This defines, in each chart, a function of $p$ mutliplied by $d\xi(p)$, i.e. locally a 1-form. 
However, nothing guarantees that it is meromorphic (even outside of isolated points), and even just continuous on some open sets.
It is easy to find some $\gamma$ such that $<\gamma,B>$ is nowhere holomorphic, as we shall see in section \ref{sec:notcycle}.
However if $\gamma$ is such that $<\gamma,B>$ is locally meromorphic in each chart, then it is independent of the choice of local coordinate $\xi$, and is globally a meromorphic 1-form on $\curve$.

\bd[Generalized cycles]
We define the space of generalized cycles as the following subspace of the algebraic dual:
\beq
\mathfrak M_1(\spcurve)
:= \{ \gamma\in \mathfrak M^1(\curve)^* \ | \ <\gamma,B>_{\pi_1} \in \mathfrak M^1(\curve)_{\pi_2} \}.
\eeq
where $\pi_1$ (resp. $\pi_2$) means the projection to the first (resp. second) factor of $\curve\times\curve$ for $B$.
Since we pair the first factor of $B$, the result is a 1-form in the second factor.
What we require here is that $<\gamma,B>_{\pi_1}$ is meromorphic in the sense of def~\ref{def:M1forms}.

$\mathfrak M_1(\spcurve)$ is a complex vector space, and we shall see below that it is of infinite uncountable dimension.

\ed

\br The space of generalized cycles is denoted $\mathfrak M_1(\spcurve)$ and not $\mathfrak M_1(\curve)$ because it depends also on $B$, so it depends on the full spectral curve, not only the underlying curve.
\er

\bd[Map $\hat B$: cycles to forms]
We define
\bea
\hat B: & \mathfrak M_1(\spcurve)  & \to  \mathfrak M^1(\curve) \cr
& \gamma & \mapsto \hat B(\gamma) = \int_\gamma B = <\gamma,B>_{\pi_1}  \ .
\eea
It is a linear map.
\ed

As we shall see below, this map is not  injective, it has a very large kernel.

\bd[Symplectic structure]
We define the symplectic form on $\mathfrak M_1(\spcurve)\times \mathfrak M_1(\spcurve)$:
\beq
\gamma_1\cap\gamma_2 := \frac{1}{2\pi\ii} \left( \int_{\gamma_1}\hat B(\gamma_2) - \int_{\gamma_2}\hat B(\gamma_1)\right).
\eeq
\ed
It obviously satisfies $\gamma_1\cap\gamma_2=-\gamma_2\cap\gamma_1$.
We will show below that $\cap$ is a non-degenerate symplectic form and thus $\mathfrak M_1(\spcurve)$ is  a symplectic space.

\bl
If $\curve = \curve_1\sqcup \curve_2$ is a disjoint union, 
if we denote the restricted spectral curves $\spcurve_i=(\curve_i,\x |_{\curve_i},\y |_{\curve_i}, B |_{\curve_i\times \curve_i})$ for $i=1,2$.
We have
\beq
\mathfrak M_1(\spcurve) = \mathfrak M_1(\spcurve_1)\oplus \mathfrak M_1(\spcurve_2).
\eeq
\el

Which means that from now on we may consider only connected spectral curves.

\proof{
By the bijection $\omega \mapsto (\omega |_{\curve_1},\omega |_{\curve_2})$ we have
\beq
\mathfrak M^1(\curve) = \mathfrak M^1(\curve_1)\oplus \mathfrak M^1(\curve_2) 
\qquad , \qquad
\mathfrak M^1(\curve)^* = \mathfrak M^1(\curve_1)^*\oplus \mathfrak M^1(\curve_2)^*.
\eeq
Let $\gamma=\gamma_1+\gamma_2\in \mathfrak M^1(\curve)^* $.
We have $\hat B(\gamma) = \omega = \omega_1+\omega_2$, where
\beq
\omega_1 = \hat B_1(\gamma_1)  + \left.\hat B(\gamma_2)\right|_{\curve_1}
\eeq
Remark that the restriction $B_{2,1}$ of $B$ to $\curve_2\times \curve_1$ is holomorphic, it has no pole, therefore $\hat B_{2,1}(\gamma_2)$ is always holomorphic on $\curve_1$.
This implies that $\hat B(\gamma)$ is meromorphic if and only if each $\hat B_i(\gamma_i)$ is meromorphic.
This implies that 
\beq
\gamma=\gamma_1+\gamma_2 \in \mathfrak M_1(\spcurve)
\quad \text{iff} \quad
\gamma_i\in \mathfrak M_1(\spcurve_i) \text{ for }i=1,2.
\eeq
}

\subsection{Explicit construction of \texorpdfstring{$\mathfrak M_1(\spcurve)$}{Lg}}

In fact we can build explicitely generating families of $\mathfrak M_1(\spcurve)$. This is done in \cite{E2017} and in fact was already in \cite{EO07}.
Let us assume that $\curve$ is connected.

\subsubsection{What is NOT a generalized cycle}
\label{sec:notcycle}

First, observe that the dual $\mathfrak M^1(\spcurve)^*$ is a very wild space, and in most cases, for an arbitrary $\gamma\in \mathfrak M^1(\spcurve)^*$, $\hat B(\gamma)$ would {\em{not}} be meromorphic, and in fact not even continuous, it can even be analytic nowhere on $\curve$.
Let us show it on an example.
Let $\gamma$ a Jordan arc, and $f$ a complex function analytic in a tubular neighborhood of $\gamma$.
The map
\bea
\td\gamma : & \mathfrak M^1(\curve) & \to \CC \cr
& \omega & \mapsto  \int_\gamma f \omega
\eea
is called a current.
One finds that $\Omega = \hat B(\td\gamma)$ is a 1-form, that is discontinuous across $\gamma$, with discontinuity:
\beq
\Omega(z^{left}) - \Omega(z^{right}) = 2\pi\ii \ df(z).
\eeq
Therefore a current can't be a generalized cycle unless $f$ is constant.

More generally if $\mu$ is a distribution on a compact closed surface $\curve$, then the following 2-dimensional integral
\beq
\hat\mu: \omega \mapsto \int_{\curve} \omega \wedge \bar d\mu
\eeq
is an element of the dual $\mathfrak M^1(\curve)^*$, but we have the Poisson relation
\beq
\bar\partial \hat B(\hat\mu) = 2\pi\ii \Delta\mu.
\eeq
In other words unless $\mu$ is harmonic outside isolated points, $\hat \mu$ cant't be a generalized cycle.

\smallskip

So now let us focus on what are generalized cycles.

\subsubsection{Canonical local coordinates}

The ramified cover $\x:\curve\to\curverond$ provides  canonical coordinates on $\curve$.

\bd[Canonical local coordinate on $\curve$]
Let $p\in\curve$. 

\begin{itemize}

\item Case $\curverond=\CC P^1$:

Let:
\begin{itemize}
\item if $\x(p)\neq \infty$: 
\beq
d_p :=  \order_p (\x-\x(p))
\qquad , \quad
x_p := \x(p).
\eeq
We have $d_p>0$.

\item if $\x(p)= \infty$: 
\beq
d_p :=  \order_p \x
\qquad , \quad
x_p := 0.
\eeq
We have $d_p<0$.

\end{itemize}

Then we define in a neighborhood of $p$:
\beq
\zeta_p(z) := (\x(z)-x_p)^{1/d_p}.
\eeq
It is a local coordinate in a neighborhood of $p$, vanishing at $p$, called the "canonical local coordinate at $p$".

If $p$ is ramified then $|d_p|>1$, and the canonical local coordinate at $p$ is defined only up to a $d_p$th root of unity, we also have the rotated canonical local coordinates 
\beq
\zeta_p(z) e^{2\pi\ii \frac{k}{d_p}}
\qquad , \ \ k=0,\dots,|d_p|-1.
\eeq

\item Case $\curverond$ a Riemann surface equiped with charts and local coordinates $\curverond\to\CC$ chosen once for all, i.e. $x\in \curverond$ can be considered in a given chart as a fixed once for all complex coordinate.
We then use for each $p$ as above, the local coordinate chart associated to $\x(p)$.

\end{itemize}

\ed

\subsubsection{Second kind cycles }

\bd[Generalized A-cycles]
Let $p\in\curve$ and an integer $k\geq 0$.
We define $\acycle_{p,k}\in \mathfrak M^1(\curve)^*$ by its pairing with any meromorphic 1-form $\omega$:
\beq
<\acycle_{p,k},\omega> := 2\pi\ii \Res_p {\zeta_p}^k \ \omega.
\eeq
$\acycle_{p,k}$ is linear in $\omega$, and is thus a well defined element of $\mathfrak M^1(\curve)^*$.
For $q\neq p$, $B(z,q)$ has no pole at $z=p$, and if $q=p$, it has a double pole at $z=p$, which gives
\beq
\hat B(\acycle_{p,k})(q) = 2\pi\ii \ \delta_{p,q} \delta_{k,1} d\zeta_p(q).
\eeq
This is a 1-form equal to $0$ except at $p$, and bounded by a rational function (in fact a constant) at $q=p$, therefore it belongs to $\mathfrak M^1(\curve)$ in the sense of def~\ref{def:M1forms}, and is in the equivalence class of $0$, i.e. $\hat B(\acycle_{p,k})\equiv 0$.
This shows that $\acycle_{p,k}\in \mathfrak M_1(\spcurve)$.
Moreover 
\beq
\acycle_{p,k}\in \Ker \hat B \subset \mathfrak M_1(\spcurve).
\eeq

\ed
(this shows that $\Ker\hat B$ is of infinite uncountable dimension).

\bd[Generalized B-cycles]
Let $p\in\curve$ and an integer $k> 0$.
We define $\bcycle_{p,k}\in \mathfrak M^1(\curve)^*$ by its pairing with any meromorphic 1-form $\omega$:
\beq
<\bcycle_{p,k},\omega> = \frac{1}{k} \Res_p \zeta_p^{-k} \ \omega.
\eeq
It is a well defined element of $\mathfrak M^1(\curve)^*$.
Then 
\beq
\hat B(\bcycle_{p,k}) = \Omega_{p,k} = \text{meromorphic 1-form that has a pole at }p\text{ of degree } k+1
\eeq
and thus belongs to $\mathfrak M^1(\curve)$,
more precisely:
\beq\label{eq:Omegapkbehav}
\Omega_{p,k} \sim \zeta_p^{-k-1}d\zeta_p + \text{holomorphic}.
\eeq
This shows that
\beq
\bcycle_{p,k}\in \mathfrak M_1(\spcurve).
\eeq

\ed
(this shows that a complement of $\Ker\hat B$ is of infinite uncountable dimension).

\bd[2nd kind times (generalized periods)]
Let $p\in\curve$ and an integer $k\geq 0$.
We define for any $\omega\in \mathfrak M^1(\curve)$:
\beq
t_{p,k}(\omega) = \frac{1}{2\pi\ii} \int_{\acycle_{p,k}}\omega = \Res_{p} \zeta_p^k \ \omega.
\eeq
Notice that 
\beq
t_{p,0}(\omega) = \Res_p \omega.
\eeq
Under a change of root of unity for the canonical local coordinate $\zeta_p\to \zeta_p e^{2\pi\ii \frac{j}{d_p}}$, we have
\beq
t_{p,k}(\omega) \to t_{p,k}(\omega) e^{2\pi\ii \frac{k j}{d_p}}.
\eeq
\ed

\bl 
For any $\omega\in \mathfrak M^1(\curve)$, 
for any $p\in\curve$, in a neighborhood of $p$ the following 1-form:
\beq
\omega - \sum_{k=0}^{\deg_p\omega -1} t_{p,k}(\omega) \ \zeta_p^{-k-1} d\zeta_p
\eeq
is holomorphic in a neighborhood of $p$ (coincides with a holomorphic 1-form except maybe at isolated points).

\el
\proof{Obvious.}

\bt[Symplectic normalization of 2nd kind $\acycle$ and $\bcycle$ cycles]
\beq
\acycle_{p,k} \cap \bcycle_{q,j} = \delta_{p,q} \delta_{k,j}.
\eeq
\et
\proof{
Easy computation using \eqref{eq:Omegapkbehav}.}

\bd[Evaluation cycle]
Let us define
\beq
\operatorname{ev}_z = d\zeta_z(z) \otimes \bcycle_{z,1}.
\eeq
If $\omega\in \mathfrak M^1(\curve) $ is holomorphic at $z$ we have
\beq
<\operatorname{ev}_z,\omega> = \omega(z)
\eeq
whence the name "evaluation" cycle.

If $\omega$ is not holomorphic at $z$, then  $<\operatorname{ev}_z,\omega>$ is the finite part of $\omega(z)$ i.e. obtained after subtracting the polar part:
\beq
<\operatorname{ev}_z,\omega> = \lim_{z'\to z, \ z'\neq z} \left( \omega(z')-\sum_{k=0}^{\deg_z\omega-1} t_{z,k}(\omega) \zeta_z(z')^{-k-1}d\zeta_{z}(z') \right).
\eeq
We have
\beq
\operatorname{ev}\in H_0(\curve,K_\curve \otimes \mathfrak M_1(\spcurve)).
\eeq

\ed

\subsubsection{Third kind cycles, Jordan arcs }

We shall immerse Jordan arcs as generalized cycles. 

Let $\mathcal J$ the set of all Jordan arcs on $\curve$, piecewise $C^1$ and with compact support and nowhere stationary (having a non-zero left and right tangent vector at every point).
Let $\gamma\in \mathcal J$ a Jordan arc, i.e. a continuous piecewise $C^1$ function with compact image,  $\gamma:I\to \curve$ where $I=[a,b] \subset \RR $ is an  interval of $\RR$ with $b>a$ (and possibly $a=-\infty$ and/or $b=+\infty$).
We denote the endpoints $p=\gamma(a)$ and $q=\gamma(b)$.
(if $p=q$ then $\gamma$ is called a Jordan loop). The image $\supp\gamma=\gamma(I)\subset \curve$ is called the support  of $\gamma$.
We also allow the point, i.e. the case $a=b$ and $\supp\gamma=\{p\}$.

\bd[Regularized integrals]
Let $\gamma\in \mathcal J$ a Jordan arc.
We define a map $\hat\gamma:\mathfrak M^1(\curve) \to \CC$ as follows:

For any $\omega\in \mathfrak M^1(\curve)$ we define
\begin{itemize}

\item if $\supp\gamma$ is a point, we define $\hat\gamma=0$, i.e.
\beq\label{eqdef:regintsupp0}
<\hat\gamma,\omega> = 0.
\eeq

\item if $\omega$ has no pole on the support of $\gamma$, we define  \beq\label{eqdef:regint0}
<\hat\gamma,\omega> = \int_\gamma \omega.
\eeq

\item if $\omega$ has a pole at the starting point $p=\gamma(a)$ and at no other $\gamma(t)$ for $t\in ]a,b]$, we let $c\in ]a,b]$ close enough to $a$ such that $\gamma|_{[a,c]}$ is in a neighborhood of $p$ in which the canonical local coordinate $\zeta_p$ is well defined. 
Then define $o=\gamma(c)$ and
\bea\label{eqdef:regint1}
<\hat\gamma,\omega> 
&=& \int_{\gamma|_{]c,b]}} \omega + \int_{\gamma|_{[a,c]}}(\omega - \sum_{k} t_{p,k}(\omega) \zeta_p^{-k-1}d\zeta_p) \cr
&& - \sum_{k=1}^{\deg_p \omega-1} \frac{t_{p,k}(\omega)}{k} \zeta_p(o)^{-k} + t_{p,0}(\omega)\left(-\ii\theta_p +\ln{\zeta_p(o)}\right)
\eea
where the branchcut of the log is defined so that $\Im\ln{\zeta_p(o)} = \operatorname{Arg}\zeta_p(o) \in [\theta_p-\pi,\theta_p+\pi[$ for $o$ close enough to $p$, and where $\theta_p = \frac{1}{d_p} \operatorname{Arg}\gamma'(a)$ is the angle of the tangent of $\gamma$ at $p=\gamma(a)$ in the coordinate $\zeta_p$.

This definition is independent of the choice of $c\in ]a,b]$, i.e. of $o\in\supp\gamma $.

\item if $\omega$ has a pole at the endpoint $q=\gamma(b)$ and at no other $\gamma(t)$ for $t\in [a,b[$, we define
\beq\label{eqdef:regintminus}
<\hat\gamma,\omega> = - <\widehat{-\gamma},\omega>
\eeq
where $-\gamma(t):=\gamma(b+a-t)$, and we use \eqref{eqdef:regint1} for $\widehat{-\gamma}$.

\item if $\omega$ has a pole at $p=\gamma(a)$ and $q=\gamma(b)$ and at no other $\gamma(t)$ for $t\in ]a,b[$, we let any $c\in ]a,b[$  and define
\beq\label{eqdef:regint2}
<\hat\gamma,\omega> = <\hat\gamma|_{[a,c]},\omega> + <\hat\gamma|_{[c,b]},\omega>
\eeq
(this is independent of the choice of $c\in ]a,b[$).

\item since $\supp\gamma$ is compact and $\omega$ meromorphic, $\omega$ can have only finitely many poles on $\supp\gamma$. 
If $\omega$ has poles at $p_1=\gamma(t_1),\dots, p_\ell=\gamma(t_\ell)$ with $t_1<t_2<\dots<t_\ell$ on the support of $\gamma$, we let $t_0=a$ and $t_{\ell+1}=b$, and define
\beq\label{eqdef:regintell}
<\hat\gamma,\omega> = \sum_{i=0}^\ell <\hat\gamma|_{[t_i,t_{i+1}]},\omega> .
\eeq

\end{itemize}
\ed

\bl[Avoiding poles from left or right]
\label{lem:avoidingpoleslr}

Let $U$ a connected compact lens tubular neighborhood of $\gamma$, such that $U\setminus\gamma$ is a union of 2 connected components $U\setminus\gamma=U_{left} \cup U_{right}$.
Let $\gamma_{left}$ (resp. $\gamma_{right}$) denote a Jordan arc homotopic deformation of $\gamma$ in $U_{left}$ (resp. in $U_{right}$), and crepant to $\gamma$ at the extremities (same tangents at the extremities).
We define
\bea
<\hat\gamma_{left},\omega> = \lim_{\text{width of } U\to 0} <\hat\gamma_{left},\omega> \cr
\qquad (\text{resp. }
<\hat\gamma_{right},\omega> = \lim_{\text{width of } U\to 0} <\hat\gamma_{right},\omega>
).
\eea
The limit is a  well defined projective limit because $<\hat\gamma_{left},\omega>$ (resp. $<\hat\gamma_{right},\omega>$) is constant when $\text{width of }U<\inf_{r=\text{poles of }\omega\text{ in }U} D(r,\gamma) $ for any distance $D$ in $U$.

We have:
\bea
\hat \gamma 
&=& \hat \gamma_{left} + \sum_{p\in \overset{\circ}{\supp\gamma}}  \frac{\pi+\theta_{p_-}-\theta_{p_+}}{2\pi} \acycle_{p,0} \cr
&=& \hat \gamma_{right} -\sum_{p\in \overset{\circ}{\supp\gamma}}\frac{\pi- \theta_{p_-}+\theta_{p_+}}{2\pi} \acycle_{p,0} \cr
&=& \frac12 \left(\hat \gamma_{left}+ \hat \gamma_{right} \right)
\eea
where, if $p=\gamma(t)$, we define $p_-$ (resp. $p_+$) = $lim_{s\to t+ 0} \gamma(s)$   (resp. = $lim_{s\to t- 0} \gamma(s)$), and $\theta_{p_-}$ (resp. $\theta_{p_+}$) is the angle (in the coordinate $\zeta_p$) of the tangent vector of $\gamma$ to the left (resp. right) of $p$.

Notice that if $p$ is a point where $\gamma$ is $C^1$, the 2 angles coincide $\theta_{p_-}=\theta_{p_+}$.

\el

\proof{
First remark that the sum $\sum_{p\in \overset{\circ}{\supp\gamma}}  c_p \acycle_{p,0} $ is a well defined generalized cycle, because  for any meromorphic $\omega$, there can be only finitely many poles on $\supp\gamma$, and 
\beq
<\sum_{p\in \overset{\circ}{\supp\gamma}}  c_p \acycle_{p,0},\omega>
= 2\pi\ii \sum_{p=\text{poles of }\omega\text{ on } \overset{\circ}{\supp\gamma}} c_p  \Res_p \omega
\eeq 
is well defined.

If $\omega$ has no pole between $p$ and $q$, the lemma is obvious.

It suffices to prove it in the case (other cases follow by concatenation) where $\omega$ has a unique pole at $r$ on $\gamma$ and no other pole, and assuming that $\gamma$ is contained in a chart where coordinate $\zeta_r$ is well defined and where $\omega$ has no other pole than $r$.

In that chart we have
\bea
<\hat\gamma,\omega>
&=& \int_{p}^r(\omega - \sum_{k=0}^{\deg_r \omega-1} t_{r,k}(\omega) \zeta_r^{-k-1}d\zeta_r) \cr
&& + \sum_{k=1}^{\deg_r \omega-1} \frac{t_{r,k}(\omega)}{k} \zeta_r(p)^{-k} - t_{r,0}(\omega)(\ln_-{\zeta_{r}(p)} -\pi\ii -\ii\theta_{r_-} )\cr
&& + \int_{r}^q(\omega - \sum_{k=0}^{\deg_r \omega-1} t_{r,k}(\omega) \zeta_r^{-k-1}d\zeta_r) \cr
&& - \sum_{k=1}^{\deg_r \omega-1} \frac{t_{r,k}(\omega)}{k} \zeta_r(q)^{-k} + t_{r,0}(\omega)(\ln_+{\zeta_{r}(q)} -\ii\theta_{r_+}) \cr
\eea
where $\ln_-$ (resp. $\ln_+$) is the determination of the logarithm with cut in the opposite direction of the left (resp. right) tangent at $r$, i.e.
\beq
\Im \ln_- {\zeta_r(z)} \in [\theta_{r_-},\theta_{r_-}+2\pi[
\quad , \quad
\Im \ln_+ {\zeta_r(z)} \in [\theta_{r_+}-\pi,\theta_{r_+} +\pi[.
\eeq

Since $\omega - \sum_{k=0}^{\deg_r \omega-1} t_{r,k}(\omega) \zeta_r^{-k-1}d\zeta_r$ has no pole on the chart, we can deform the integration contour to $\gamma_{left}$:
\bea
<\hat\gamma,\omega>
&=& \int_{\gamma_{left}} (\omega - \sum_{k=0}^{\deg_r \omega-1} t_{r,k}(\omega) \zeta_r^{-k-1}d\zeta_r) \cr
&& + \sum_{k=1}^{\deg_r \omega-1} \frac{t_{r,k}(\omega)}{k} \zeta_r(p)^{-k} - t_{r,0}(\omega)\ln_-{\zeta_{r}(p)} \cr
&& - \sum_{k=1}^{\deg_r \omega-1} \frac{t_{r,k}(\omega)}{k} \zeta_r(q)^{-k} + t_{r,0}(\omega)\ln_+{\zeta_{r}(q)} \cr
&& + \ii t_{r,0}(\omega) (\pi +\theta_{r_-}-\theta_{r_+}) \cr
&=& \int_{\gamma_{left}} \omega  - \sum_{k=0}^{\deg_r \omega-1} t_{r,k}(\omega)  \int_{\gamma_{left}} \zeta_r^{-k-1}d\zeta_r) \cr
&& + \sum_{k=1}^{\deg_r \omega-1} \frac{t_{r,k}(\omega)}{k} \zeta_r(p)^{-k} - t_{r,0}(\omega)\ln_-{\zeta_{r}(p)} \cr
&& - \sum_{k=1}^{\deg_r \omega-1} \frac{t_{r,k}(\omega)}{k} \zeta_r(q)^{-k} + t_{r,0}(\omega)\ln_+{\zeta_{r}(q)} \cr
&& + \ii t_{r,0}(\omega) (\pi +\theta_{r_-}-\theta_{r_+}) \cr
&=& \int_{\gamma_{left}} \omega  
 -  t_{r,0}(\omega)  \left( \int_{\gamma_{left}} \zeta_r^{-1}d\zeta_r  +\ln_-{\zeta_{r}(p)} - \ln_+{\zeta_{r}(q)} \right) \cr
&& + \ii t_{r,0}(\omega) (\pi +\theta_{r_-}-\theta_{r_+}) \cr
\eea
The integral of $ \zeta_r^{-1}d\zeta_r$ on $\gamma_{left}$ (resp. $\gamma_{right}$) goes above (resp. below) zero, and gives $\ln_+{\zeta_{r}(q)}- \ln_-{\zeta_{r}(p)} $ modulo $2\pi\ii$, and eventually we get the result.
}

\bt[Map Jordan arcs $\to$ generalized cycles ]

We have

\begin{itemize}

\item The map $\gamma\mapsto\hat\gamma$ is well defined,
\item $<\hat\gamma,\omega>$ is a linear in $\omega$, i.e. $\hat\gamma\in \mathfrak M^1(\curve)^*$.

\item $\hat\gamma$ is invariant by reparametrization of $\gamma$, i.e. if $f:[a',b']\to [a,b]$ is a monotonic piecewise $C^1$ bijection, we have
\beq
\widehat{\gamma\circ f} = \hat\gamma.
\eeq

\item The map $\gamma\mapsto\hat\gamma$ is additive under concatenation, and satisfies
\beq
\widehat{\gamma_1+\gamma_2} = \hat{\gamma_1} + \hat{\gamma_2} \qquad , \quad \widehat{-\gamma}=-\hat\gamma.
\eeq

\item $\hat\gamma$ is a generalized cycle, $\hat\gamma\in \mathfrak M_1(\spcurve)$.
This defines a map than immerses Jordan arcs into generalized cycles:
\bea
\mathcal J & \to & \mathfrak M_1(\spcurve) \cr
\gamma & \mapsto & \hat\gamma
\eea
which is such that for any meromorphic 1-form $\omega\in \mathfrak M^1(\curve)$ not having pole on the support of $\gamma$ we have
\beq
<\hat\gamma,\omega> = \int_{\hat\gamma} \omega = \int_\gamma \omega.
\eeq
(this justifies the Poincarr\'e pairing notation and the name "generalized cycles").

\item $\hat B(\hat\gamma)$ is a meromorphic 1- form that has poles at the extremities of $\gamma$, a simple pole at $p$ of residue $-1$, a simple pole at $q$ of residue $+1$ and no other pole on $\curve$ (whence the name 3rd kind cycle):
\beq
-\Res_p \hat B(\hat\gamma) = \Res_q \hat B(\hat\gamma) = 1.
\eeq

\item It can be extended to a linear map on a $\mathbb A$-module  subring of $\CC$ (like $\mathbb A =\ZZ$ or $\mathbb Q$ or $\RR$ or $\CC$  ) of linear combinations of Jordan arcs:
\beq
\widehat{\sum_{i=1}^k c_i \gamma_i} = \sum_{i=1}^k c_i \hat{\gamma_i}.
\eeq

\end{itemize}

\et

\br[Identifying $\gamma$ with $\hat\gamma$]
From now on, thanks to this map, and for the sake of simplifying notations, we shall identify a Jordan arc $\gamma$ with its generalized cycle $\hat\gamma$.
This is what is done most of the time in the literature of topological recursion.
\er

\proof{

\begin{itemize}

\item First the map $\hat\gamma$ is well defined. One can check that it is independent of $o$ in \eqref{eqdef:regint1} (take the derivative wrt $o$), independent of $c$ in \eqref{eqdef:regint2}. Each summand in \eqref{eqdef:regintell} is exclusively defined by \eqref{eqdef:regint1} or \eqref{eqdef:regint2} or \eqref{eqdef:regintminus} or \eqref{eqdef:regintsupp0} or \eqref{eqdef:regint0}, so is well defined.

\item It is easy to see that each term is linear in $\omega$, so that $\hat\gamma\in \mathfrak M^1(\curve)^*$.

\item $\hat\gamma$ is invariant by reparametrization because all terms are.

\item It is easy to see that $\hat\gamma$ is additive by concatenation, nearly from construction (it is defined by adding concatenations of pieces).

\item To prove that $\hat\gamma$ is a generalized cycle, we need to compute $\Omega_\gamma=\hat B(\hat\gamma)$.

Let $z\in\curve$
\begin{itemize}
\item if $z\notin \supp\gamma$, then $B(.,z)$ has no pole on $\supp\gamma$, and $\Omega_\gamma(z)=\int_\gamma B(.,z)$ is a convergent integral, and locally analytic of $z$.

\item if $z\in \overset{\circ}{\supp\gamma}$,  such that $z\neq p$ and $z\neq q$, lemma \ref{lem:avoidingpoleslr} gives
\bea
\Omega_\gamma(z) 
&=& <\hat\gamma_{left},B(.,z)> + \pi\ii \sum_{r\in \overset{\circ}{\supp\gamma}} \Res_r B(.,z) \cr
&=& <\hat\gamma_{left},B(.,z)> + \pi\ii  \Res_z B(.,z) \cr
\eea
Since $B$ is symmetric, it has a double pole at $z$ without residue, so that $\Res_z B(.,z)=0$.

Moreover, since $z\notin\gamma_{left}$ we are back to the 1st situation, and $\Omega_{\gamma}$ is locally analytic of $z$.
%

\item If $z$ is in a neighborhood of $p$ and outside of $\supp\gamma$, we have
\bea
\Omega_\gamma(z) 
&=& \int_{z'=p}^o B(z',z) + \int_{z'=o}^q B(z',z) \cr
&=& \int_{z'=p}^o \frac{d\zeta_p(z')\otimes d\zeta_p(z)}{(\zeta_p(z)-\zeta_p(z'))^2} \cr
&& + \int_{z'=p}^o \left(B(z',z)-\frac{d\zeta_p(z')\otimes d\zeta_p(z)}{(\zeta_p(z)-\zeta_p(z'))^2} \right) + \int_{z'=o}^q B(z',z) \cr
&=& - \frac{d\zeta_p(z)}{\zeta_p(z)} + \frac{d\zeta_p(z)}{\zeta_p(z)-\zeta_p(o)} \cr
&& + \int_{z'=p}^o \left(B(z',z)-\frac{d\zeta_p(z')\otimes d\zeta_p(z)}{(\zeta_p(z)-\zeta_p(z'))^2} \right) + \int_{z'=o}^q B(z',z) \cr
&=& - \frac{d\zeta_p(z)}{\zeta_p(z)} + \text{holomorphic at }p.
\eea

\item similarly is $z$ is in a neighborhood of $q$ we have
\bea
\Omega_\gamma(z) 
&=& \frac{d\zeta_q(z)}{\zeta_p(z)} + \text{holomorphic at }q.
\eea

\end{itemize}

Putting all together we obtain that $\Omega_{\gamma} = \hat B(\hat\gamma)$ is a meromorphic 1-form with a simple pole at $p=\gamma(a)$ with residue $-1$, a simple pole at $q=\gamma(b)$ with residue $+1$, and no other pole. And if $p=q$ it is a holomorphic form without any poles on $\curve$.
This implies that $\hat\gamma\in \mathfrak M_1(\spcurve)$.

\end{itemize}

}

\bt[Intersections of Jordan arcs]
\label{th:interArcs}
Let $\gamma_1$ and $\gamma_2$ two Jordan arcs.

\begin{itemize}

\item  if $\gamma_1$ and $\gamma_2$ don't intersect we have
\beq
\hat\gamma_1 \cap \hat\gamma_2 =  0 = \gamma_1 \cap \gamma_2 .
\eeq
where $\gamma_1\cap\gamma_2 $ is the usual intersection form. 

\item if $\gamma_1$ and $\gamma_2$ intersect  transversally at a point $p$ different from the extremities, and the tangent of $\gamma_i$ is in the direction of angle $\theta_i$, and $\theta_2-\theta_1\in ]-\pi,\pi[$, we have
we have
\beq
\hat\gamma_1 \cap \hat\gamma_2 = \operatorname{sign}(\theta_2-\theta_1)  = \gamma_1 \cap \gamma_2.
\eeq

\item
if the intersection $p$ is the starting point of $\gamma_1$ and a middle point of $\gamma_2$ we have
\beq
\hat\gamma_1 \cap \hat\gamma_2 = \frac12 \operatorname{sign}(\theta_2-\theta_1)  .
\eeq

\item
if the intersection is the starting point of $\gamma_1$ and the starting point of $\gamma_2$ we have
\beq
\hat\gamma_1 \cap \hat\gamma_2 = \frac12 \operatorname{sign}(\theta_2-\theta_1) - \frac{1}{2\pi}(\theta_2-\theta_1)  \in [-\frac12,\frac12].
\eeq

\item if $p$ is an ending point, revert $\gamma\to -\gamma$ and use that
\beq
(-\hat\gamma_1)\cap \hat\gamma_2 = -\hat\gamma_1\cap \hat\gamma_2
\quad , \quad
\hat\gamma_2\cap\hat\gamma_1 = -\hat\gamma_1\cap\hat\gamma_2.
\eeq

\item if there are several intersection points, the intersection is the algebraic sum of contributions of each intersection points.

\end{itemize}

\et

\proof{


Let $\gamma_1:[a,b]\to \curve$ and $\gamma_2:[\td a,\td b]\to\curve$.
We have to compute
\beq
\int_{z\in\gamma_1} \left( \int_{z'\in\gamma_2} B(z',z)\right)  - \int_{z'\in\gamma_2} \left( \int_{z\in\gamma_2} B(z',z)\right) .
\eeq

$\bullet$ If $\gamma_1$ and $\gamma_2$ do not intersect, the pole of $B$ at $z=z'$ is never on the integration paths, and the order of integration commutes, and the difference gives 0.

$\bullet$ Now consider the case where $\gamma_1$ and $\gamma_2$ have an intersection point $\gamma_1\cap\gamma_2 =\{p\}$, which we assume transverse.
Let $r>0$, small enough  such that inside the disc $D(p,r)$ of center $p$ and radius $r$,  $\Gamma_1 = \gamma_1\cap D(p,r)  $ is a connected  Jordan arc and $\Gamma_2 = \gamma_2\cap D(p,r)$ is a connected Jordan arc.
Let us decompose the paths $\gamma_i$ with the part outside the disc $D(p,r)$ and the part inside the disc:
\bea
\gamma_1 = \bar\gamma_1 + \Gamma_1
\quad , \quad
\gamma_2 = \bar\gamma_2 + \Gamma_2.
\eea
We thus have
\bea
\int_{z\in\gamma_1} \left( \int_{z'\in\gamma_2} B(z',z)\right) 
&=& \int_{z\in\bar\gamma_1} \left( \int_{z'\in\bar\gamma_2} B(z',z)\right)   
 + \int_{z\in\bar\gamma_1} \left( \int_{z'\in\Gamma_2} B(z',z)\right)   \cr
&& + \int_{z\in\Gamma_1} \left( \int_{z'\in\bar\gamma_2} B(z',z)\right)   
 + \int_{z\in\Gamma_1} \left( \int_{z'\in\Gamma_2} B(z',z)\right)   \cr
\eea
In the first 3 integrals the pole of $B(z,z')$ is not on the integration paths and thus the order of integration commutes.
This implies that we can reduce the computation only for the parts inside the disc:
\beq
2\pi\ii \hat\gamma_1\cap\hat\gamma_2 = 2\pi\ii \hat\Gamma_1\cap\hat\Gamma_2.
\eeq
Inside the disc $D(p,r)$ we use the local coordinate $z=\zeta_p$ and write
\beq
B(z',z) = \frac{dz'\otimes dz}{(z'-z)^2} + f(z',z) \ dz' \otimes dz
\eeq
where $f(z',z)$ is analytic in $D(p,r)\times D(p,r) $.
In particular the double integral with $f$ behaves as $O(r^2)$ as $r\to 0$.

Let us assume that the tangents of $\gamma_1$ and $\gamma_2$ at $p$ are in direction $\theta_1$ and $\theta_2$, and assume $\theta_2-\theta_1 \in ]-\pi,\pi[$.
Let us define the following integral, that is the integral from $p$ on the half-Jordan arcs starting from $p$ of $\gamma_1$ and $\gamma_2$:
\bea
I_{\theta_1,\theta_2} 
&=& \Im \int_{\e^{\ii\theta_1}0_+}^{\e^{\ii\theta_1}r } dx \int_{\e^{\ii\theta_2} 0_+}^{\e^{\ii\theta_2} r} \frac{dx'}{(x-x')^2} \cr
&=& \Im \int_{0_+}^{1} dx  \left( \frac{1}{x- e^{\ii(\theta_2-\theta_1)}  } - \frac{1}{x} \right) \cr
&=& \operatorname{Arg} \frac{1- e^{\ii(\theta_2-\theta_1)} }{- e^{\ii(\theta_2-\theta_1)} } \cr
&=& \operatorname{Arg} (1-e^{-\ii(\theta_2-\theta_1)} ) \cr
&=& \frac12 (\pi\operatorname{sign}(\theta_2-\theta_1) - (\theta_2-\theta_1)) \ \ \in [-\pi/2,\pi/2].
\eea
The integrals computing the intersection are then sums taking into account all ways to glue the half-Jordan arcs to make the full arcs.

$\bullet$ If $p$ is not an endpoint for both paths, we have:
\bea
&& (I_{\theta_1,\theta_2} - I_{\theta_1,\theta_2\pm\pi}) - (I_{\theta_1\pm\pi,\theta_2} - I_{\theta_1\pm\pi,\theta_2\pm\pi}) \cr
&=& (\frac12(\pi+\theta_1-\theta_2) - \frac12( \theta_1-\theta_2)) - (\frac12(\theta_1-\theta_2) - \frac12(\pi+\theta_1-\theta_2)) \cr
&=& \pi \operatorname{sign}(\theta_2-\theta_1)
\eea
This gives
\bea
2\pi \hat\gamma_1\cap \hat\gamma_2  
&=& \left( (I_{\theta_1,\theta_2} - I_{\theta_1,\theta_2\pm\pi})-(I_{\theta_1\pm\pi,\theta_2} - I_{\theta_1\pm\pi,\theta_2\pm\pi}) \right) \cr
&& - \left( (I_{\theta_2,\theta_1} - I_{\theta_2,\theta_1\pm\pi}) -(I_{\theta_2\pm\pi,\theta_1} - I_{\theta_2\pm\pi,\theta_1\pm\pi}) \right)  \cr
&=& \pi(\operatorname{sign}(\theta_2-\theta_1)-\operatorname{sign}(\theta_1-\theta_2)) 
\eea
i.e.
\beq
\hat\gamma_1\cap \hat\gamma_2 = \operatorname{sign}(\theta_2-\theta_1) = \gamma_1\cap \gamma_2.
\eeq

$\bullet$ Case $p$ is the starting point of $\gamma_1$ and a middle point of $\gamma_2$, we have

\bea
2\pi \hat\gamma_1\cap \hat\gamma_2  
&=& \left( I_{\theta_1,\theta_2} - I_{\theta_1,\theta_2\pm\pi} \right) \cr
&& - \left( I_{\theta_2,\theta_1}  -I_{\theta_2\pm\pi,\theta_1} \right)  \cr
&=& \pi \operatorname{sign}(\theta_2-\theta_1)
\eea
i.e.
\beq
\hat\gamma_1\cap \hat\gamma_2 = \frac12 \operatorname{sign}(\theta_2-\theta_1) .
\eeq

$\bullet$ Case $p$ is the starting point of $\gamma_1$ and of $\gamma_2$, we have
\bea
2\pi \hat\gamma_1\cap \hat\gamma_2  
&=&  I_{\theta_1,\theta_2}  -  I_{\theta_2,\theta_1}  \cr
&=& \pi\operatorname{sign}(\theta_2-\theta_1)- (\theta_1-\theta_1)
\eea
i.e.
\beq
\hat\gamma_1\cap \hat\gamma_2 = \frac12 \operatorname{sign}(\theta_2-\theta_1)- \frac{1}{2\pi} (\theta_2-\theta_1) .
\eeq
Notice that
\beq
\hat\gamma_1\cap \hat\gamma_2 \in [-\frac12,\frac12].
\eeq

If instead of being the starting point $p$ is an ending point, we just revert $\gamma_i\to -\gamma_i$,
and if there are several intersection points we take the algebraic sum.

}

\subsubsection{First kind cycles }

First kind cycles are the images of closed Jordan loops, those verifying $\gamma(a)=\gamma(b)$.
\bd[First kind cycles]
Let 
\beq
\mathcal J_0 = \{ \gamma \in \mathcal J \ | \ \gamma(a)=\gamma(b)\}.
\eeq
We have a map
\bea
\mathcal J_0 & \to &  \mathfrak M_1(\spcurve)\cr
\gamma & \mapsto & \hat\gamma.
\eea
It can be extended to a linear map on the module of linear combinations of Jordan loops:
\beq
\widehat{\sum_{i=1}^k c_i \gamma_i} = \sum_{i=1}^k c_i \hat{\gamma_i}.
\eeq

\ed

The following theorem is obvious
\bt
If $\gamma$ is a Jordan loop,
then
\beq
\hat B(\hat\gamma) = \text{holomorphic 1-form on }\curve \in \Omega^1(\curve).
\eeq

\et

As a corollary of theorem~\ref{th:interArcs} we have
\bt[Intersections]
The intersection of 1st kind generalized cycles is the same as the usual intersection of the corresponding homology classes of Jordan loops
\beq
\hat\gamma_1\cap\hat\gamma_2=  \gamma_1\cap\gamma_2.
\eeq

\et

\subsubsection{Boundary cycles}

Assume that $\curve$ has some boundary $b$ that is topologicaly a circle, oriented so that $\curve$ is to the right of $b$, and such that $\x(b)=b_0$ is a circle on the base $\curverond$, and we choose a local coordinate on $\curverond$ so that the circle $\x(b)$ is the unit circle $S^1 \subset \CC$, and the image of a neighborhood of $b$ is sent to the exterior of the unit disc in $\CC$.

Notice that $\x$ being a ramified covering map, it may happen that $\x$ in a neighborhood of $b$ is not 1:1, but $d_b:1$.
We say that $d_b\geq 1 $ is the \textbf{winding number} of $b$.

We define
\bd[Local coordinate on $b$]
\beq
\zeta_b(p) := \x(p)^{1/d_b}.
\eeq
The coordinate $\zeta_b$ is well defined in an annulus $1<|\zeta_b|<r_b$.
\ed

Also we recall that our requirement for meromorphic 1-forms in that they are meromorphic on the interior of $\curve$, which means that they can be singular on the boundary.

\bd[Boundary cycles]
We associate a family of generalized cycles to a boundary $b$ as follows:

\begin{itemize}

\item Cycle $\acycle_{b,k}$. For $k\geq 0$, we define
\bea
\acycle_{b,k}: & \mathfrak M^1(\curve) & \to \CC \cr
& \omega & \mapsto <\acycle_{b,k},\omega> = \liminf_{r \to 1+} \int_{\x^{-1}(r S^1) } \zeta_b^k \ \omega = \liminf_{r \to 1+} \int_{\theta=0}^{2\pi} r^k \ e^{\ii k \theta} \ \omega. \cr
\eea
The $\liminf$ is a well defined projective limit, because the integral is independent of $r$ if $1<r<r_\omega$ with $r_\omega \leq  \min\{|\zeta_b(p)| \ | \ p=\text{poles of }\omega \}\cup\{r_b\}$. 

\item Cycle $\bcycle_{b,k}$. For $k>0$ we define
\bea
\bcycle_{b,k}: & \mathfrak M^1(\curve) & \to \CC \cr
& \omega & \mapsto <\bcycle_{b,k},\omega> = \frac{1}{2\pi\ii k} \liminf_{r \to 1+} \int_{\x^{-1}(r S^1) } \zeta_b^{-k} \ \omega.
\eea
Again, the $\liminf$ is a well defined projective limit, because the integral is independent of $r$ if $1<r<r_\omega$.

\item Cycle $\bcycle_{b,0}$. 
For $k=0$, let $o$ some generic point of $\curve$ and $\gamma_{o\to{\zeta_{b_j}^{-1}(r)}}$ a Jordan arc from $o$ to the point of coordinate $\zeta_{b_j}=r$ on the boundary $b_j$. 
We define
\bea
\bcycle_{b,0,o}: & \mathfrak M^1(\curve) & \to \CC \cr
& \omega & \mapsto <\bcycle_{b,0,o},\omega> = 
\liminf_{r \to 1+} \Big( \int_{\hat\gamma_{o\to{\zeta_{b_j}^{-1}(r)}}} \omega + \frac{-1}{2\pi}  \int_{\theta=0}^{2\pi } \theta \ \omega(\zeta_b^{-1}(r e^{\ii \theta})) \cr
&& \qquad \qquad \qquad \qquad - \frac{\ln r}{2\pi\ii}\int_{\acycle_{b_j,0}} \omega \Big).
\eea
The $\liminf$ is a well defined projective limit because the right hand side is independent of $r$ for $1<r<r_\omega$.

\end{itemize}

\ed

\bt
$\acycle_{b,k}$ and $\bcycle_{b,k}$ are generalized cycles.

We have
\beq
\acycle_{b,k} \cap \bcycle_{b',k'} = \delta_{b,b'} \delta_{k,k'}.
\eeq
\et
\proof{The fact that the $\liminf$ exist because these integrals are independent of $r$ if $1<r<r_\omega$ implies that the pairing to any meromorphic 1-form is a complex number, so these are linear forms on $\mathfrak M^1(\curve)$. The fact that integrating $B(.,z_0)$ yields a meromorphic 1-form is again because if $1<r<|\zeta_b(z_0)|$, the integration path doesn't contain the pole of $B(.,z_0)$.}

\br[Poles as boundaries]
One can remove a small disc around a pole and make it a boundary, and then these definitions coincide with the definitions of 2nd kind cycles associated to poles.
This remark is very useful to do surgery of surfaces.
\er

\bl[Fourier series]

Any $\omega\in \mathfrak M^1(\curve)$ is holomorphic in the tubular neighborhood of $b$ the annulus $1<|\zeta_b|<r_\omega$.
In the local coordinate $\zeta_b$ we can write
\beq
\omega = f(\zeta_b) d\zeta_b = f(r e^{\ii\theta}) (dr + \ii r  d\theta)e^{\ii\theta} 
\quad
\text{where }  \zeta_b = r e^{\ii\theta} \ , \ r_\omega>r>1.
\eeq
The function $f(r e^{\ii\theta})$ is a periodic $C^\infty$ function of $\theta$ of period $2\pi $, for any $r$ in a neighborhood of $1$.
It can thus be decomposed as a Fourier series:
\beq
\label{eq:Fourierseriesb}
\om(z) = \sum_{k=-\infty}^\infty c_{b,k} \zeta_b(z)^{-k-1}d\zeta_b(z) 
\qquad , \quad
c_{b,k} = \frac{r^{k}}{2\pi\ii  } \int_{\theta=0}^{2\pi} e^{k\ii\theta} \omega
= \frac{1}{2\pi\ii  } \int_{\theta=0}^{2\pi} \zeta_b^{k} \omega
\eeq
The Fourier series is absolutely and uniformly convergent for any $r_\omega>r>1$, which implies that at large $k>0$ we have
\beq
|c_{b,k}|=O(r_\omega^k)
\qquad , \qquad
|c_{b,-k}| = O(1).
\eeq
Notice that if $k\geq 0$:
\beq
<\acycle_{b,k},\omega> = \int_{\acycle_{b,k}} \omega = 2\pi\ii c_{b,k} = O(r_\omega^k) \ \ \text{for large }k\geq 0,
\eeq
and if $k>0$
\beq
<\bcycle_{b,k},\omega> = \int_{\bcycle_{b,k}} \omega = \frac{1}{k } c_{b,-k} = O(1) \ \ \text{for large }k> 0.
\eeq

\el
\proof{Fourier series of periodic $C^\infty$ functions are absolutely  convergent.}

\subsection{Inverse map: forms to cycles}

In this section Let us assume that $\curve$ is an open  connected surface of some genus $\genus$ with $n\geq 0$ boundaries that are topological circles denoted $b_i=\partial_i \curve$, $i=1,\dots,n$. And if $n=0$, $\curve$ is compact of genus $\genus$ without boundary.

Let us also add the requirement that $B$ is normalized on boundaries, i.e. for each boundary $b_j$:
\beq
\label{assump:Bbndaries}
\hat B(\acycle_{b_j,0})=0.
\eeq

\bd[Marking]
Choose a generic interior point $o\in \curve$.
Then we choose $n$ non-intersecting $C^1$ Jordan arcs $\gamma_{o\to b_j}$ that end at the boundary $b_j=\partial_j \curve$ at the point of coordinate $\zeta_{b_j}=1$.
Then $\pi_1(\curve\setminus \cup_{j=1}^n \gamma_{o\to b_j})$ has rank $2\genus$, so we can choose a generating family of $2\genus$ non-intersecting Jordan loops starting end ending at $o$, denoted $\acycle_1,\dots,\acycle_{\genus}, \bcycle_1,\dots,\bcycle_{\genus}$, that do not intersect each other except at $o$, and with symplectic intersections
\beq
\acycle_i\cap \bcycle_j = \delta_{i,j}
\quad , \quad
\acycle_i\cap \acycle_j = 0
\quad , \quad
\bcycle_i\cap \bcycle_j = 0.
\eeq
Such a choice of $2\genus+n$ Jordan arcs is not unique, we choose one of the possible, and call it a "torelli marking of Jordan loops".
\ed

\bd[Fundamental domain]
Having chosen a Torelli marking of loops, we define the graph $\Upsilon$ as the union of all arcs
\beq
\Upsilon = \cup_{j=1}^{\genus} \acycle_j\cup_{j=1}^{\genus} \bcycle_j \cup_{j=1}^{n} \gamma_{o\to b_j} \cup_{j=1}^{n} b_j
\eeq
and the fundamental domain
\beq
\curve_0 = \curve\setminus \Upsilon.
\eeq
By our definition, $\curve_0$ is simply connected.

The boundaries of $\curve_0$ is
\bea
\label{eqdef:bndFundDomain}
\partial \curve_0
&=  &
\sum_{j=1}^{\genus} \acycle_{j}^{right} - \acycle_{j}^{left}  \cr
&& + \sum_{j=1}^{\genus} \bcycle_{j}^{right} - \bcycle_{j}^{left}  \cr
&& + \sum_{j=1}^{n} \gamma_{o\to b_j}^{right} -\gamma_{o\to b_j}^{left} \cr
&& + \sum_{j=1}^{n} \zeta_{b_j}^{-1}(S_1) .
\eea

\ed
%
%

\bd[Fundamental bi-cycle]

Having chosen a fundamental domain $\curve_0=\curve\setminus \Upsilon$, we define:
\bea
2\pi\ii \ \hat \Gamma
&=& \sum_{i=1}^{\genus}  \hat\acycle_i \otimes \hat\bcycle_i-\hat\bcycle_i\otimes \hat\acycle_i \cr
&& + \sum_{p\in\curve}\sum_{k=1}^\infty \acycle_{p,k} \otimes \bcycle_{p,k} - \bcycle_{p,k} \otimes \acycle_{p,k} \cr
&& + \sum_{p\in\curve} \acycle_{p,0} \otimes \hat\gamma_{o\to p} - \hat\gamma_{o\to p} \otimes \acycle_{p,0}  \cr
&& + \sum_{j=1}^n \sum_{k=0}^\infty \acycle_{b_j,k} \otimes \bcycle_{b_j,k} - \bcycle_{b_j,k} \otimes \acycle_{b_j,k} \cr
\eea
where, for each $p\in\curve$,  $\gamma_{o\to p}$ is a once for all chosen Jordan arc from $o$ to $p$ in the fundamental domain. 
An example is to choose a Jordan arc geodesic for the distance $|\y| $ on $\curve$ (there is typically only finitely many geodesics from $o$ to $p$).
\ed
It may look like an ill-defined uncountably infinite sum of tensor product of generalized cycles. However, as we shall see below, it is a well defined element of $\mathfrak M^1(\curve)^*\otimes \mathfrak M^1(\curve)^*$, because when paired with meromorphic forms $\omega_1\otimes \omega_2$, it reduces to finite or absolutely convergent sums.

\bl
\beq
\hat\Gamma \in \mathfrak M_1(\spcurve)\otimes \mathfrak M_1(\spcurve).
\eeq

\el

\proof{
Let $\omega_1\otimes \omega_2 \in \mathfrak M^1(\curve)\otimes \mathfrak M^1(\curve) $ .
We have
\bea
<\hat\Gamma , \omega_1\otimes \omega_2>
&=& \frac{1}{2\pi\ii}\sum_{i=1}^{\genus}  \int_{\hat\acycle_i}\omega_1 \int_{\hat\bcycle_i} \omega_2 - \int_{\hat\bcycle_i}\omega_1 \int_{\hat\acycle_i} \omega_2 \cr
&& + \sum_{p=\text{poles of }\omega_1} \sum_{k=1}^{\deg_p \omega_1-1} t_{p,k}(\omega_1) \int_{\bcycle_{p,k}} \omega_2 \cr
&& - \sum_{p=\text{poles of }\omega_2} \sum_{k=1}^{\deg_p \omega_2-1} t_{p,k}( \omega_2) \int_{\bcycle_{p,k}}\omega_1  \cr
&& + \sum_{p=\text{poles of }\omega_1}  \Res_p \omega_1 \  \int_{\hat\gamma_{o\to p}} \omega_2 \cr
&& - \sum_{p=\text{poles of }\omega_2}  \int_{\hat\gamma_{o\to p}}\omega_1 \ \Res_p \omega_2 \cr
&& + \frac{1}{2\pi\ii}\sum_{i=1}^n \sum_{k=0}^\infty \int_{\acycle_{b_i,k}}\omega_1 \int_{\bcycle_{b_i,k}}\omega_2 - \int_{\bcycle_{b_i,k}} \omega_1 \int_{\acycle_{b_i,k}}\omega_2 \cr
\eea
The term in the last line is an absolutely convergent infinite sum, and all the others are  finite sums.
This implies that $\hat\Gamma$ is a well defined element of $\mathfrak M^1(\curve)^*\otimes \mathfrak M^1(\curve)^* $.
Since it involves only combinations of generalized cycles, it is also in $\mathfrak M_1(\spcurve)\otimes \mathfrak M_1(\spcurve) $.
}

\bt[Riemann Bilinear indentity]
\label{prop:RBI}
If $\omega_1,\omega_2 \in \mathfrak M^1(\curve) $, we have
\beq
<\hat\Gamma,\omega_1\otimes \omega_2>=0.
\eeq 

\et

\textbf{Proof:}
In appendix~\ref{App:proofRBI}.

\bc
\label{th:RBIhatB}

It satisfies
\beq
<\hat\Gamma,\omega \otimes B(.,z_0)> = \omega(z_0)
\eeq

\ec

\proof{
Let $z_0\in\curve$ in a chart with some local coordinate $\zeta(z_0)$.
Let
\beq
\omega_2(z) = B(z,z_0) / d\zeta(z_0)
\eeq
it is a meromorphic 1-form $\omega_2\in \mathfrak M^1(\curve)$, with a double pole at $z=z_0$ and no other pole, and near the pole it behaves like
\beq
\omega_2(z) \sim  \frac{1}{(\zeta(z)-\zeta(z_0))^2}d\zeta(z) + \text{holomorphic}
\eeq
and in particular it has $t_{z_0,1}(\omega_2)=1$ and $t_{z_0,0}(\omega_2)=0$.
We then apply \autoref{prop:RBI} to $\omega\otimes \omega_2$ and, putting on one side all the terms coming from the poles of $\omega_2$ i.e. from $z_0$, we have:
\beq
d\zeta(z_0) \ \sum_{p=\text{poles of }\omega_2} \int_{\acycle_{p,k}}\omega_2 \int_{\bcycle_{p,k}}\omega
= 2\pi\ii d\zeta(z_0)  \ \int_{\bcycle_{z_0,1}}\omega = 2\pi\ii <\operatorname{ev}_{z_0},\omega>  = 2\pi\ii \omega(z_0).
\eeq
All the other terms applied to $\omega_2$ give the left hand side of:
\beq
<\hat\Gamma,\omega \otimes B(.,z_0)> = \omega(z_0).
\eeq
}

\bd[Operator $\hat C$: forms to cycles]
Let
\bea
\hat C: & \mathfrak M^1(\curve)  & \to   \mathfrak M_1(\spcurve)  \cr
& \omega & \mapsto  \hat \Gamma(\omega,.) 
\eea

\ed

\br
Heuristically, if there would exist a basis $\{\gamma_i\}_{i\in J}$ of $\mathfrak M_1(\spcurve)$, with intersection matrix $I_{i,j}=\gamma_i\cap \gamma_j$ invertible, we would define
\beq
\hat C(\omega) = \sum_{i,j\in J^2} \frac{1}{2\pi\ii} \left(\int_{\gamma_i}\omega\right) \ (I^{-1})_{i,j} \ \gamma_j.
\eeq
However such a basis doesn't exist.
But what the definition of $\hat C$ above does, thanks to Riemann bilinear identity, is to provide an adapted basis for each $\omega$, where the sums are actually finite or absolutely convergent.

\er

\bt
$\hat B$ is surjective
and $\hat C$ is a right inverse of $\hat B$:
\beq
\hat B \circ \hat C = \text{Id}.
\eeq
\et

\proof{
Follows from  corollary~\ref{th:RBIhatB}.

}

\subsubsection{Lagrangian decomposition}

\bt[Lagrangian decomposition]
We have the Lagrangian decomposition 
\beq
\mathfrak M_1(\spcurve) = \Ker\hat B \oplus \Im \hat C,
\eeq
where both $\Ker \hat B$ and $\Im \hat C$ are Lagrangian.

The map $\hat B:\mathfrak M_1(\spcurve) \to \mathfrak M^1(\curve)$ is surjective.

We have the exact sequence
\beq
0 \to \Ker \hat B \to \mathfrak M_1(\spcurve) \overset{\hat B}{\to} \mathfrak M^1(\curve) \to 0.
\eeq

\et

\proof{
It is obvious from the definitions that $\cap$ vanishes on $\Ker\hat B\times \Ker\hat B$.

Let $\gamma\in \mathfrak M_1(\spcurve)$ and $\gamma'=\hat C(\hat B(\gamma))$, we have $\gamma'\in \Im \hat C$.
And we have
\beq
\hat B(\gamma')= \hat B\circ \hat C(\hat B(\gamma)) = \hat B(\gamma)
\eeq
therefore $\hat B(\gamma-\gamma')=0$ which implies $\gamma-\gamma'\in \Ker\hat B$.
This shows that
\beq
\mathfrak M_1(\spcurve) = \Ker\hat B \oplus \Im \hat C.
\eeq

It remains to prove that $\hat C$ is Lagrangian.
We have
\bea
2\pi\ii \hat C(\omega_1) \cap \hat C(\omega_2)
&=& \int_{\hat C(\omega_1)} \hat B(\hat C(\omega_2)) - \int_{\hat C(\omega_2)} \hat B(\hat C(\omega_1)) \cr
&=& \int_{\hat C(\omega_1)} \omega_2 - \int_{\hat C(\omega_2)} \omega_1 \cr
&=& <\hat\Gamma,\omega_1\otimes \omega_2> - <\hat\Gamma,\omega_2\otimes \omega_1> \cr
&=& 2  \ <\hat\Gamma,\omega_1\otimes \omega_2> 
\eea
which vanishes due to \autoref{prop:RBI}.
}

\bt
The Intersection symplectic form is non degenerate.

\et

\proof{
Assume that there would exist some $\gamma\in \mathfrak M_1(\spcurve)$, such that $\forall \td\gamma\in \mathfrak M_1(\spcurve)$ we would have $\gamma\cap\td\gamma=0$.
In particular this would imply
\bea
\int_\gamma \hat B(\td\gamma) = \int_{\td\gamma} \hat B(\gamma).
\eea
In particular, with $\td\gamma = \hat C(\omega)$, we would have
\beq
\int_\gamma \omega = 
\int_\gamma \hat B\circ \hat C(\omega) = \int_{\hat C(\omega)} \hat B(\gamma) = <\hat\Gamma,\omega\otimes \hat B(\gamma)> = 0
\eeq
This would imply that for any $\omega$ we would have $\int_\gamma\omega=0$.
In other words, as an element of $\mathfrak M^1(\curve)$ we would have $\gamma=0$.

This implies that $\cap$ is non degenerate.
}

%
%

\bd[Generating cycle of  $\Ker \hat B$]

Let
\beq
\acycle(z_0) = \hat C(B(.,z_0)).
\eeq
More explicitely:
\bea
\acycle(z_0)
&=& \sum_{i=1}^{\genus} \left( \int_{\bcycle_i} B(.,z_0) \right) \ \acycle_i - \left( \int_{\acycle_i} B(.,z_0) \right) \ \bcycle_i  \cr
&& + \sum_{j=1}^{n} \sum_{k=1}^\infty \left( \int_{\bcycle_{b_j,k}} B(.,z_0) \right) \ \acycle_{b_j,k} 
 - \sum_{j=1}^{n} \sum_{k=1}^\infty \left( \int_{\acycle_{b_j,k}} B(.,z_0) \right) \ \bcycle_{b_j,k} \cr
&& + \sum_{j=1}^{n} \left( \int_{\bcycle_{b_j,0}} B(.,z_0) \right) \ \acycle_{a_j,0} - \left( \int_{\acycle_{b_j,0}} B(.,z_0) \right) \ \bcycle_{a_j,0} .
\eea
It is a 1-form in $z_0$, valued in generalized cycles,
it belongs to
\beq
H^0(\curve, K_\curve\otimes \mathfrak M_1(\spcurve)).
\eeq

\ed

\bt
For every $z_0\in \curve$, $\acycle(z_0)$ is in $\Ker\hat B$ (in fact $\acycle(z_0)\in T^*_{z_0}\curve \otimes \Ker\hat B$):
\beq
\forall z_0\in \curve, \ z_1\in \curve, \  
\qquad
\int_{\acycle(z_0)} B(.,z_1) = 0.
\eeq
Vice versa, $\Ker \hat B$ is generated by $\acycle(z_0)$, i.e. for every $\gamma\in \Ker\hat B$ we have
\beq
\gamma = \int_{z_0\in\gamma} \acycle(z_0) .
\eeq


\et

\proof{}

We have:
\beq
<\acycle(z_0),B(.,z_1)> = 
<\hat\Gamma, B(.,z_0)\otimes B(.,z_1)>
= B(z_0,z_1)-B(z_1,z_0) = 0 .
\eeq

Then, we have
\beq
2\pi\ii   \acycle(z_0)\cap \td\gamma
= \int_{\acycle(z_0)} \hat B(\td\gamma) - \int_{\td\gamma} \hat B(\acycle(z_0)) = \int_{\acycle(z_0)} \hat B(\td\gamma)  .
\eeq

Let $\gamma\in \Ker\hat B$.
For every $\omega\in \mathfrak M^1(\curve)$ we have
\beq
<\int_{z_0\in\gamma} \acycle(z_0) , \omega >
= \int_\gamma \int_{\hat C(\omega)} B 
= \int_\gamma \hat B(\hat C(\omega)) 
= \int_\gamma \omega.
\eeq
This implies that, as elements of $\mathfrak M^1(\curve)^*$ we have
\beq
\gamma = \int_{z_0\in\gamma} \acycle(z_0) .
\eeq

{}


\section{Topological Recursion}

The main application framework of the generalized cycles is Topological Recursion.

\subsection{Topological Recursion}

Topological recursion \cite{EO07} is a recursive procedure that associates a sequence of forms to a spectral curve
\bea
& \text{Topological Recursion (TR)} & \cr
\spcurve & \mapsto & \{ \omega_{g,n}(\spcurve)\}_{(g,n) \in \ZZ_+\times \ZZ_+ \setminus \{(0,0)\} }
\eea
where $\omega_{0,1}=\y$, $\omega_{0,2}=B$, and 
for $2g-2+n>0$ and $n>0$
\beq
\omega_{g,n} \in H^0(\curve^n , K_\curve^{\overset{\text{sym}}{\boxtimes} n}(*R))
\eeq
where $R$ is the set of ramification points of $\x$.

For $n=0$, $\omega_{g,0}$ is often denoted $F_g$ and is a scalar $\in \CC$:
\beq
\omega_{g,0}(\spcurve)=F_g(\spcurve).
\eeq

Moreover we have the following useful lemma:
\bl[Normalized on $\Ker\hat B$]
\label{lem:omgnKerB}
\beq
\forall \gamma\in \Ker\hat B, \ \forall g\geq 0 , \ n\geq 1+\delta_{g,0} \qquad \quad
<\gamma, \omega_{g,n}>_{\pi_1} =0.
\eeq

\el
\proof{
See \cite{EO07}. Notice that the case $(g,n)=(0,2)$ is the definition of $\Ker\hat B$.
}

\subsection{Deformations of Topological Recursion}

In \cite{EO07} there is the following theorem:
\bt[Deformations \cite{EO07}]
Let $\spcurve_t$ for $t\in I=[a,b]\subset \RR$ a smooth family of spectral curves.
If one can find some generalized cycle $\Gamma_t\in \mathfrak M_1(\spcurve_t)$ such that:
\beq
\frac{d}{dt}\y_t = \int_{\Gamma_t} B_t
\eeq
\beq
\frac{d}{dt}B_t = \int_{\Gamma_t} \omega_{0,3}(\spcurve_t)
\eeq
then one has the special geometry relation for all $(g,n)\neq (0,0)$:
\beq
\frac{d}{dt}\omega_{g,n}(\spcurve_t) = \int_{\Gamma_t} \omega_{g,n+1}(\spcurve_t).
\eeq
\et

The following lemma follows easily from lemma \ref{lem:omgnKerB}
\bl
The genearlized cycle $\Gamma_t$ is not unique, it can be defined modulo $\Ker\hat B_t$.
This allows to choose $\Gamma_t$ in some Lagrangian submanifold $\mathcal L_t\subset \mathfrak M_1(\spcurve_t)$, transverse to $\Ker \hat B_t$.
\el

Let us make the following definition:
\bd[flat cycles]
The family $\Gamma_t$ is said "constant" or "flat" iff:
\begin{itemize}
\item it is continuous, i.e. $\Gamma_t$ is continuous in any finite dimensional sub-bundle of $\mathfrak M^1(\curve_t)\to I$,

\item we have
\beq
\frac{d}{dt}\x_{t*}\Gamma_t=0
\eeq
where the push-forward $\x_{t*}\Gamma_t \in \mathfrak M^1(\curverond)^* $ is defined by $<\x_{t*}\Gamma_t,\Omega> = <\Gamma_t,\x_t^*\Omega>$. 

In fact flat families of cycles can be used to define a connection of the bundle of generalized cycles.
\end{itemize}

\ed

\bt
If $\Gamma_t$ is constant, we can exponentiate the deformation:
\beq
\omega_{g,n}(\spcurve_{t+h}) = \sum_{m=0}^\infty \frac{h^m}{m!} \overbrace{\int_{\Gamma_t}\dots\int_{\Gamma_t}}^{m} \omega_{g,n+m}(\spcurve_t)
\eeq
The sum over $m$  is absolutely convergent for $h$ small enough.

\et
\proof{This is Taylor series.}

\subsection{Prepotential}

We can understand why $F_0=\omega_{0,0}$ is ill defined.
Consider 2 flows $d/dt$ and $d/d\td t$, corresponding to 2 generalyzed cycles $\Gamma$ and $\td\Gamma$, assumed constant.

If $\omega_{0,0}$ would exist we would have 
\beq
\frac{d}{dt}\omega_{0,0} = \int_{\Gamma} \y
\qquad , \qquad
\frac{d}{d\td t}\omega_{0,0} = \int_{\td\Gamma} \y
\eeq
and taking a second derivative
\beq
\frac{d}{d\td t}\frac{d}{d t}\omega_{0,0}- \frac{d}{d t}\frac{d}{d \td t}\omega_{0,0} = \int_{\Gamma}\int_{\td\Gamma} B -  \int_{\td\Gamma}\int_{\Gamma} B = 2\pi \ii \ \Gamma\cap \td\Gamma.
\eeq
It would vanish iff $\Gamma\cap \td\Gamma=0$, i.e. iff $\Gamma$ and $\td\Gamma$ are in a same flat Lagrangian $\mathcal L$.

In other words, we can define $\omega_{0,0}$ only if we also chose a flat Lagrangian $\mathcal L$.

We have from \cite{E2017}:
\bd[Prepotential]
Let $\spcurve$ a spectral curve, and $\mathcal L$  a flat Lagrangian.
We define:
\beq
F_0(\spcurve,\mathcal L)
=\frac12 \left< \Pi^{\parallel\mathcal L}_{\Ker\hat B}\circ\hat C(\y),\y\right> 
\eeq
where $\Pi^{\parallel\Ker\hat B}_{\mathcal L}:\mathfrak M_1(\spcurve)\to \Ker\hat B$ is the projection onto $\mathcal L$, parallel to   $\Ker\hat B$.
\ed

\bt
For any $\Gamma_t\in\mathcal L$ associated to a family $\spcurve_t$, we have
\beq
\frac{d}{dt} F_0(\spcurve_t,\mathcal L) = \int_{\Gamma_t} \y.
\eeq
\et
\proof{In \cite{E2017}}

\bt
If we have a multidimensional family $\spcurve_{t_1,\dots,t_k}$, we let 
\beq
\delta = \sum_{i=1}^k \Gamma_{i} dt_i.
\eeq
It is a generalized cycle valued in the cotangent space of the $t_i$s, i.e. a 1-form in the cotangent moduli space, valued in generalized cycles.

We have
\beq
d F_0(\spcurve,\mathcal L) = <\delta,\y> = \sum_{i=1}^k <\Gamma_i,\y> dt_i.
\eeq
Since $\mathcal L$ is Lagrangian we have the Lax equation (zero curvature equation)
\beq
[d,d]=0 \qquad \Leftrightarrow \quad \Gamma_i\cap\Gamma_j=0.
\eeq
\et
\proof{In \cite{E2017}}

\section{Conclusion}

The aim of this article was to provide a comprehensive setting of the notion of generalized cycles.
Their geometry is fascinating, and they have plenty of applications in integrable systems, and topological recursion.

In \cite{EO07} they were introduced as a way to encode deformations.

They provide a way to unify Miwa-Jimbo equations, Seiberg-Witten and Bertola Malgrange into a single notion.

It was shown in \cite{eynard2019topological}, how to formulate Kontsevich-Soibelman quantum Airy structures \cite{kontsevich2017airy} in terms of generalized cycles. The formalism is powerful enough to easily provide the generalization of 
quantum Airy structures to higher order operators (limited to quadratic operators in Kontsevich-Soibleman).

\smallskip

Generalized cycles are also useful in understanding the relationship between the moduli space of spectral curves, and the deRham moduli space of connections on principal bundles \cite{Belliard_2018}.
%
In \cite{belliard2024Lie} we study how in the $\epsilon\to 0$ limit, the Goldman cycles project to the present generalized cycles of a spectra curve which is the Hictchin's map $\det(y-\Phi(x))$.

\section*{Acknowledgments}

This work is supported by the ERC synergy grant ERC-2018-SyG  810573, ``ReNewQuantum''.

\setcounter{section}{0}

\appendix{Riemann Bilinear identity}
\label{App:proofRBI}

\bt[Fundamental bi-cycle and Riemann Bilinear Identity]

It satisfies
\beq
<\hat\Gamma,\omega_1 \otimes \omega_2> = 0.
\eeq

\et

\proof{}
The proof uses Riemann bilinear identity, which proceeds as follows:

Let $r>0$ small enough such that for each pole $p$ of $\omega_1$ and/or $\omega_2$, the coordinate $\zeta_p$ is well defined in the disc $|\zeta_p|<r$, all these discs are disjoints, and $\omega_1$ and $\omega_2$ are analytic in the pointed disc (i.e. the only pole is at $\zeta_p=0$ in each disc), and for each boundary $b$, $\omega_1 $ and $\omega_2$ are analytic in the annulus of radius $1<|\zeta_b|<r$, and all the annulus and discs are disjoint.

Then, in the fundamental domain $\curve_0$, let us consider Jordan arcs $\gamma_{o\to p}$ for each pole $p$ of $\omega_1$ and/or $\omega_2$, and such that in the discs $|\zeta_p|<r$, the Jordan arc's support is the segment $[0,r]$ oriented from $r$ to $0$.
We define the truncated arc at raidus $r$
\beq
\gamma_{o\to p}^{(r)} = \gamma_{o\to p} \setminus \zeta_p^{-1}([0,r]).
\eeq
Let also the small circle of radius $r$ around $p$, starting at angle $\theta=0$ and ending at $\theta=2\pi$:
\beq
\mathcal C_p^{(r)} = \zeta_p^{-1}(\{r e^{\ii \theta}  \ | \ \theta\in ]0,2\pi[ \}).
\eeq

Let 
\beq
\curve_0' = \curve_0\setminus \cup_{p=\text{poles of }\omega_2}\{\gamma_{o\to p}\}
\eeq
In $\curve_0'$, let an antiderivative of $\omega_2$:
\beq
\phi_2(z) = \int_o^z \omega_2
\eeq
i.e. $d\phi_2=\omega_2$.
By definition $\phi_2$ is analytic in $\curve_0'$.

In $\curve'_0$, which is simply connected, the following contour is contractible:
\bea
\gamma
&=& \sum_{i=1}^\genus \acycle_i^{right} -\acycle_i^{left} \cr
&& +\sum_{i=1}^\genus \bcycle_i^{right} -\bcycle_i^{left} \cr
&& +\sum_{p=\text{poles of }\omega_2} \gamma_{o\to p}^{(r) right} - \gamma_{o\to p}^{(r) left} + \mathcal C_p^{(r)} \cr
&& +\sum_{j=1}^n \gamma_{o\to b_j}^{(r) right} - \gamma_{o\to b_j}^{(r) left} + \mathcal C_{b_j}^{(r)} .
\eea
It may however enclose the poles of $\omega_1$, and therefore we have:
\beq
\int_{\partial \curve_0'} \omega_1 \phi_2
= -2\pi\ii \sum_{p=\text{poles of }\omega_1} \Res_p \phi_2 \ \omega_1 
\eeq
where by "poles of $\omega_1$", we mean the poles of $\omega_1$ inside $\curve'_0$, i.e. those that are not poles of $\omega_2$.

This gives
\bea
0
&=& \sum_{p=\text{ poles of }\omega_1} 2\pi\ii \Res_{p} \phi_2 \omega_1  + \int_{\partial\curve'_0} \phi_2  \ \omega_1 \cr
&=& \sum_{p=\text{ poles of }\omega_1} 2\pi\ii \Res_{p} \phi_2 \  \omega_1 \cr
&& + \sum_{p=\text{ poles of }\omega_2} \left( \int_{\hat\gamma^{right}_{o\to p^{(r)}}} \phi_2(z) \ \omega_1(z) \right) - \left( \int_{\hat\gamma^{left}_{o\to p^{(r)}}} \phi_2(z) \ \omega_1(z) \right) \cr
&& + \sum_{p=\text{ poles of }\omega_2}   \int_{\theta=0, \ z=\zeta_p^{-1}(r e^{\ii\theta})}^{2\pi} \phi_2(z) \ \omega_1(z)  \  \cr
&& + \sum_{i=1}^{\genus} \left( \int_{\acycle_i^{right}} \phi_2(z) \ \omega_1(z) \right) - \left( \int_{\acycle_i^{left}} \phi_2(z) \ \omega_1(z) \right) \cr
&& + \sum_{i=1}^{\genus} \left( \int_{\bcycle_i^{right}} \phi_2(z) \ \omega_1(z) \right) - \left( \int_{\bcycle_i^{left}} \phi_2(z) \ \omega_1(z) \right) \cr
&& + \sum_{j=1}^{n} \left( \int_{\hat\gamma^{right}_{o\to p^{(r)}_{b_j}}} \phi_2(z) \ \omega_1(z) \right) - \left( \int_{\hat\gamma^{left}_{o\to p^{(r)}_{b_j}}} \phi_2(z) \ \omega_1(z) \right) \cr
&& + \sum_{j=1}^{n}   \int_{\theta=0, \ z=\zeta_b^{-1}(r e^{\ii\theta})}^{2\pi} \phi_2(z) \ \omega_1(z)  \  \cr
\eea

Let us compute each term:
\begin{itemize}

\item
Let $p$ a pole of $\omega_1$ which is not a pole of $\omega_2$, we have:
\bea
&& \Res_{z\to p} \phi_2(z) \omega_1(z) \cr
&=& \Res_{z\to p} \phi_2(z) \left(\sum_{k=0}^{\deg_p \omega_1-1} t_{p,k}(\omega_1) \zeta_p(z)^{-k-1}d\zeta_p(z) + \text{hol. at }p \right)  \cr
&=& \sum_{k=1}^{\deg_p \omega_1-1} t_{p,k}(\omega_1) \Res_{z\to p}  \phi_2(z) \zeta_p(z)^{-k-1}d\zeta_p(z)   \cr
&& + t_{p,0}(\omega_1) \Res_{z\to p}  \phi_2(z) \zeta_p(z)^{-1}d\zeta_p(z)   \cr
&=& \sum_{k=1}^{\deg_p \omega_1-1} \frac{1}{k} t_{p,k}(\omega_1) \Res_{z\to p}  \omega_2(z) \zeta_p(z)^{-k} + t_{p,0}(\omega_1) \ \phi_2(p) \cr
&=& \sum_{k=1}^{\deg_p \omega_1-1}  t_{p,k}(\omega_1)  \int_{\bcycle_{p,k}}\omega_2 + t_{p,0}(\omega_1) \ \int_{\hat\gamma_{o\to p}} \omega_2 \cr
&=& 2\pi\ii \sum_{k=1}^{\deg_p \omega_1-1}  \int_{\acycle_{p,k}}\omega_1 \int_{\bcycle_{p,k}}\omega_2 + 2\pi\ii \int_{\acycle_{p,0}}\omega_1  \ \int_{\hat\gamma_{o\to p}} \omega_2 \cr
\eea
Notice that $\int_{\hat\gamma_{o\to p}} \omega_2$ is independent of the choice of Jordan arc $\gamma_{o\to p}$ in $\curve'_0$, because $\curve'_0$ is simply connected and  $\omega_2$ is analytic in $\curve'_0$.

Notice also that $\int_{\acycle_{p,k}} \omega_2=0$ so we can also write
\bea
&& \Res_{z\to p} \phi_2(z) \omega_1(z) \cr
&=& 2\pi\ii \sum_{k=1}^{\deg_p \omega_1-1}  \left( \int_{\acycle_{p,k}}\omega_1 \int_{\bcycle_{p,k}}\omega_2 
- \int_{\bcycle_{p,k}}\omega_1 \int_{\acycle_{p,k}}\omega_2 \right) \cr
&& + 2\pi\ii \left( \int_{\acycle_{p,0}}\omega_1  \ \int_{\hat\gamma_{o\to p}} \omega_2 
- \ \int_{\hat\gamma_{o\to p}} \omega_1 \int_{\acycle_{p,0}}\omega_2  \right) .
\eea

\item For the 2nd term, let $p$ a pole of $\omega_2$. 
Notice that for a pole $p$ of $\omega_2$, $\phi_2$ is possibly discontinuous across $\gamma_{o\to p}$, with dicontinuity
\beq
\phi^{right}_2(z) - \phi^{left}_2(z)
= 2\pi\ii t_{p,0}(\omega_2) = \int_{\acycle_{p,0}}\omega_2
\eeq
which is independent of $z$.
This gives
\bea
&& \left( \int_{\hat\gamma^{right}_{o\to p^{(r)}}} \phi_2(z) \ \omega_1(z) \right) - \left( \int_{\hat\gamma^{left}_{o\to p^{(r)}}} \phi_2(z) \ \omega_1(z) \right) \cr
&=& -2\pi\ii \ t_{p,0}(\omega_2) \int_{\hat\gamma_{o\to p^{(r)}}} \omega_1 \cr
&=& - \int_{\acycle_{p,0}} \omega_2 \int_{\hat\gamma_{o\to p^{(r)}}} \omega_1 \cr
\eea

\item For $p$ a pole of $\omega_2$, we use that:
\beq
\omega_1(z) \sim \sum_{k=0}^{\deg_p\omega_1-1} t_{p,k}(\omega_1) \ \zeta_p(z)^{-k-1} d\zeta_p(z)
+ \sum_{k=1}^{\infty} k \int_{\bcycle_{p,k}} \omega_1 \ \  \zeta_p(z)^{k-1} d\zeta_p(z)
\eeq
and
\bea
\phi_2(z) 
%
&\sim & \int_{\hat\gamma_{o\to p}}\omega_2 + t_{p,0}(\omega_2) (-\pi\ii+\ln\zeta_p(z))  - \sum_{k=1}^{\deg_p\omega_2-1} \frac{1}{k}t_{p,k}(\omega_2) \zeta_p(z)^{-k} \cr
&&  + \sum_{k=1}^{\infty}\zeta_p(z)^{k} \int_{\bcycle_{p,k}}\omega_2 \   \cr
\eea
This gives
\bea
&&  \int_{\theta=0, \ z=\zeta_p^{-1}(r e^{\ii\theta})}^{2\pi} ( \phi_2(z) \ \omega_1(z)  \  \cr
&=& 2\pi\ii \left(\int_{\hat\gamma_{o\to p}}\omega_2 -\pi \ii t_{p,0}(\omega_2) \right) \ t_{p,0}(\omega_1) \cr
&& - 2\pi\ii \sum_{k=1}^{\deg_p\omega_2-1} t_{p,k}(\omega_2) \ \int_{\bcycle_{p,k}}\omega_1 \cr
&& + 2\pi\ii \sum_{k=1}^{\deg_p\omega_1-1} t_{p,k}(\omega_1) \ \int_{\bcycle_{p,k}}\omega_2 \cr
&& + t_{p,0}(\omega_2) \int_{\theta=0, \ z=\zeta_p^{-1}(r e^{\ii\theta})}^{2\pi} \ln\zeta_p(z) \ \omega_1(z) \cr
&=&  2\pi^2 t_{p,0}(\omega_2) \ t_{p,0}(\omega_1) \cr
&& + \int_{\hat\gamma_{o\to p}}\omega_2 \int_{\acycle_{p,0}}\omega_1
 +  \sum_{k=1}^{\deg_p\omega_1-1} \int_{\acycle_{p,k}}\omega_1 \ \int_{\bcycle_{p,k}}\omega_2 - \int_{\bcycle_{p,k}}\omega_1 \ \int_{\acycle_{p,k}}\omega_2 \cr
&& + t_{p,0}(\omega_2) \int_{\theta=0, \ z=\zeta_p^{-1}(r e^{\ii\theta})}^{2\pi} \ln\zeta_p(z) \ \omega_1(z) \cr
\eea
For the last term we consider $\phi_1$ an antiderivative of $\omega_1$, such that $d\phi_1=\omega_1$, namely
\bea
\phi_1(z) 
&= & \int_{\hat\gamma_{o\to p}}\omega_1 + t_{p,0}(\omega_1) (-\pi\ii+\ln\zeta_p(z))  - \sum_{k=1}^{\deg_p\omega_1-1} \frac{1}{k}t_{p,k}(\omega_1) \zeta_p(z)^{-k} \cr
&&  + \sum_{k=1}^{\infty} \zeta_p(z)^{k}  \int_{\bcycle_{p,k}}\omega_1 \  .
\eea
We have
\bea
&& \int_{\theta=0, \ z=\zeta_p^{-1}(r e^{\ii\theta})}^{2\pi} \ln\zeta_p(z) \ \omega_1(z)  \cr
&=& \left(\int_{\hat\gamma_{o\to p}}\omega_1 -\pi\ii t_{p,0}(\omega_1) \right) 2\pi\ii  \cr
&& + \frac12 t_{p,0}(\omega_1) ( -4\pi^2 + 4\pi\ii \ln r) \cr
&& - 2\pi\ii \sum_{k=1}^{\deg_p\omega_1-1} \frac{r^{-k}}{k} t_{p,k}(\omega_1)   + 2\pi\ii \sum_{k=1}^{\infty} r^k \int_{\bcycle_{p,k}}\omega_1 \cr
&=& 2\pi\ii \phi_1(\zeta_p^{-1}(r)) + \pi\ii t_{p,0}(\omega_1)
\eea
This gives
\bea
&&  \int_{\theta=0, \ z=\zeta_p^{-1}(r e^{\ii\theta})}^{2\pi}  \phi_2(z) \ \omega_1(z)  \cr
&& + \left( \int_{\hat\gamma^{right}_{o\to p^{(r)}}} \phi_2(z) \ \omega_1(z) \right) - \left( \int_{\hat\gamma^{left}_{o\to p^{(r)}}} \phi_2(z) \ \omega_1(z) \right) \cr
&=&   \int_{\hat\gamma_{o\to p}}\omega_2 \int_{\acycle_{p,0}}\omega_1
 +  \sum_{k=1}^{\deg_p\omega_1-1} \int_{\acycle_{p,k}}\omega_1 \ \int_{\bcycle_{p,k}}\omega_2 - \int_{\bcycle_{p,k}}\omega_1 \ \int_{\acycle_{p,k}}\omega_2 \cr
&& + \int_{\acycle_{p,0}} \omega_2 \left( - \int_{\hat\gamma_{o\to p^{(r)}}} \omega_1 + \phi_1(\zeta_p^{-1}(r))  \right) \cr
&=&   \int_{\hat\gamma_{o\to p}}\omega_2 \int_{\acycle_{p,0}}\omega_1 - \int_{\acycle_{p,0}}\omega_2 \int_{\hat\gamma_{o\to p}}\omega_1 \cr
&&  +  \sum_{k=1}^{\deg_p\omega_1-1} \int_{\acycle_{p,k}}\omega_1 \ \int_{\bcycle_{p,k}}\omega_2 - \int_{\bcycle_{p,k}}\omega_1 \ \int_{\acycle_{p,k}}\omega_2 \cr
\eea

\item $\acycle_i$ and $\bcycle_i$ cycles:
\bea
&&  \left( \int_{\acycle_i^{right}} \phi_2(z) \ \omega_1(z) \right) - \left( \int_{\acycle_i^{left}} \phi_2(z) \ \omega_1(z) \right) \cr
&=&  \int_{\acycle_i} \left(  \phi_2(z^{right}) - \phi_2(z^{left}) \right) \ \omega_1(z)  \cr
&=&  \int_{\acycle_i} \left(  \int_{\bcycle_i} \omega_2 \right) \ \omega_1(z)  \cr
&=&  \int_{\acycle_i} \omega_1 \ \int_{\bcycle_i}\omega_2
\eea
Similarly, but using that the orientation of $\bcycle_i$ compared to that of $\acycle_i$ is opposite, we have
\bea
&&  \left( \int_{\bcycle_i^{right}} \phi_2(z) \ \omega_1(z) \right) - \left( \int_{\bcycle_i^{left}} \phi_2(z) \ \omega_1(z) \right) \cr
&=& - \int_{\bcycle_i} \omega_1 \ \int_{\acycle_i}\omega_2  .
\eea
Here we used the same notation $\acycle_i$, $\bcycle_i$ for the Jordan loop and its corresponding generalized cycle, for lighter formulas.

\item Boundaries $b_j$:

The proof for boundaries is exactly the same as the proof for poles, replacing the Laurent expansion for poles, by the Fourier series, which takes the same form:
\beq
\omega_1 
\sim \sum_{k=0}^\infty \left(\frac{1}{2\pi\ii }\int_{\acycle_{b_j,k}}  \omega_1\right) \zeta_{b_j}^{-k-1}d\zeta_{b_j}
+ \sum_{k=1}^\infty \left(\int_{\bcycle_{b_j,k}}  \omega_1\right) k \zeta_{b_j}^{k-1}d\zeta_{b_j}.
\eeq

\end{itemize}

Eventually all the terms together give
\beq
<\hat\Gamma,\omega_1\otimes \omega_2>=0.
\eeq
which completes the proof.
{}

\bibliography{texte}

\end{document}